\documentclass[10pt,twocolumn,twoside,journal]{IEEEtran}

\pdfoutput=1 

\IEEEoverridecommandlockouts 

\ifCLASSINFOpdf
  \usepackage[pdftex]{graphicx}
  \graphicspath{{./figures/}}
  \DeclareGraphicsExtensions{.pdf,.jpeg,.png}
\else
  \usepackage[dvips]{graphicx}
  \graphicspath{{./figures/}}
  \DeclareGraphicsExtensions{.eps}
\fi

\usepackage[cmex10]{amsmath}
\usepackage{amsfonts,amssymb,bbm}
\usepackage{mathtools} 
\usepackage{color}
\usepackage{cite}
\usepackage{array}
\usepackage[caption=false,font=footnotesize]{subfig}
\usepackage{fixltx2e}
\usepackage{url}
\usepackage{booktabs}
\usepackage{enumerate}

\def\bsK{{\boldsymbol{K}}}

\def\calB{{\mathcal{B}}}
\def\calC{{\mathcal{C}}}

\def\calE{{\mathcal{E}}}

\def\calI{{\mathcal{I}}}
\def\calJ{{\mathcal{J}}}

\def\calN{{\mathcal{N}}}

\def\calR{{\mathcal{R}}}

\def\calT{{\mathcal{T}}}
\def\calU{{\mathcal{U}}}
\def\calV{{\mathcal{V}}}

\def\calX{{\mathcal{X}}}
\def\calY{{\mathcal{Y}}}

\def\Real{{\mathbb{R}}}
\def\Natural{{\mathbb{N}}}

\def\E{{\mathbb{E}}}

\DeclareMathAlphabet{\CMmathcal}{OMS}{cmsy}{m}{n}
\renewcommand{\mathcal}[1]{\CMmathcal{#1}}
\newcommand{\figref}[1]{Fig.~\ref{#1}}

\newtheorem{theorem}{Theorem}
\newtheorem{lemma}{Lemma}
\newtheorem{proposition}{Proposition}
\newtheorem{corollary}{Corollary}
\newtheorem{definition}{Definition}
\newtheorem{remark}{Remark}

\newcommand*{\QEDA}{\hfill\IEEEQED}%

\def\el{~et~al.\ }

\newcommand{\eqtop}[1]{\overset{(#1)}{=}}
\newcommand{\leqtop}[1]{\overset{(#1)}{\leq}}
\newcommand{\geqtop}[1]{\overset{(#1)}{\geq}}

\begin{document}

\title{The CEO Problem With Secrecy Constraints}
%
%
%
\author{Farshad~Naghibi,~\IEEEmembership{Student~Member,~IEEE},
    Somayeh~Salimi,~\IEEEmembership{Member,~IEEE},
    and~Mikael~Skoglund,~\IEEEmembership{Senior~Member,~IEEE}%
    \thanks{\copyright~2014 IEEE. Personal use of this material is permitted. However, permission to use this material for any other purposes must be obtained from the IEEE by sending a request to pubs-permissions@ieee.org}%
    \thanks{Authors are with the School of Electrical Engineering and ACCESS Linnaeus Center,
    KTH Royal Institute of Technology, SE-100 44 Stockholm, Sweden (emails: \{naghibi,somayen,skoglund\}@ee.kth.se).}%
    \thanks{Part of the material in this work was presented in the IEEE International Symposium on Information Theory (ISIT), Honolulu, HI, 2014 \cite{Naghibi-2014-ISIT}.}%
    }

\maketitle

\begin{abstract}
We study a lossy source coding problem with secrecy constraints in which a remote information source should be transmitted to a single destination via multiple agents in the presence of a passive  eavesdropper. The agents observe noisy versions of the source and independently encode and transmit their observations to the destination via noiseless rate-limited links. The destination should estimate the remote source based on the information received from the agents within a certain mean distortion threshold. The eavesdropper, with access to side information correlated to the source, is able to listen in on one of the links from the agents to the destination in order to obtain as much information as possible about the source. This problem can be viewed as the so-called CEO problem with additional secrecy constraints. We establish inner and outer bounds on the rate-distortion-equivocation region of this problem. We also  obtain the region in special cases where the bounds are tight. Furthermore, we study the quadratic Gaussian case and provide the optimal rate-distortion-equivocation region when the eavesdropper has no side information and an achievable region for a more general setup with side information at the eavesdropper.
\end{abstract}

\begin{IEEEkeywords}
CEO problem, multiterminal source coding, secrecy constraints, eavesdropping, equivocation.
\end{IEEEkeywords}

%
\IEEEpeerreviewmaketitle

\section{Introduction}

As networks are becoming more distributed, their vulnerability to malicious activities increases which in turn raises the concern on the security of such networks. Consequently, information-theoretic security as a concrete framework for analyzing secrecy in networks has gained attention among researchers \cite{Liang-2008-FTCIT,Liu_SecuringWireless}. Information-theoretic security, which was initially introduced by Shannon \cite{Shannon-1949-Bell}, exploits different statistical characteristics of received information at the legitimate receiver and at the eavesdropper. Moreover, it makes no assumptions on the computational power of the eavesdropper, unlike the traditional cryptographic approaches for secrecy. Later, Wyner introduced the Wiretap channel model in \cite{Wyner-1975-Bell} and showed that perfectly secure communication without a shared secret key is possible if the channel from the transmitter to the eavesdropper is a degraded version of the channel to the legitimate receiver. This result was generalized to broadcast channels with confidential messages by Csisz\'{a}r and K\"{o}rner in \cite{Csiszar-1978-TIT}. Subsequently, many extensions to this problem have been developed and studied in the literature (see, for instance, \cite{Liang-2008-FTCIT}, \cite{Liu_SecuringWireless}, and references therein).

In this paper, we consider secrecy in a multiterminal source coding problem. In particular, we study the problem of conveying an information source to a single destination via multiple agents (encoders) in the presence of a passive eavesdropper. The agents have access to noisy observations of the source and are connected to the destination via noiseless rate-limited links. They do not cooperate or communicate to one another and are not required to estimate the source themselves. This scenario is of interest for many applications such as sensor networks or smart grid systems where reconstruction of the source at sensors and smart meters is not necessary. The distributed nature of such networks makes them more susceptible to eavesdropping. At each instant, the eavesdropper listens in on one of the links from the agents to the destination in order to obtain information about the source. In addition, it has access to side information correlated to the source. Since the link that will be compromised by the eavesdropper is unknown to the agents prior to their transmissions, each agent should protect its link in order to leak as little information as possible about the source. Our objective is to characterize the trade-off among agents' transmission rates, incurred distortion at the destination, and the amount of information revealed to the eavesdropper. %
This setup can be viewed as the extension of the so-called CEO problem \cite{Berger-1996-TIT} in which  secrecy constraints are considered.


\subsection{Related Work}

The chief executive/estimation officer (CEO) problem was motivated in \cite{Berger-1996-TIT} by a communication and distributed processing system analogous to a scenario in which a firm's CEO is interested in  information of a source that cannot be observed directly. The CEO assigns a group of agents to independently observe a corrupted version of the source and communicate their observations. The lossless variant of this setup was initially studied by Gel'fand and Pinsker \cite{Gelfand-1979-PIT}. It was extended by Yamamoto and Itoh \cite{Yamamoto-1980-IECE} as well as Flynn and Gray \cite{Flynn-1987-TIT} to the lossy case with only two encoders for which an achievable rate-distortion region was derived. The model was generalized to the CEO problem with many encoders by Berger and Viswanathan \cite{Berger-1996-TIT} in which the trade-off between the end-to-end average distortion and sum of the rates at which the agents transmit to the CEO was studied. Multiterminal lossy source coding problems, including the CEO problem, are still open in general. However, for the special case of the quadratic Gaussian CEO problem \cite{Viswanathan-1997-TIT}, the sum-rate-distortion function for infinite number of agents with identical signal-to-noise ratios (SNRs) was derived by Oohama \cite{Oohama-1998-TIT}, and later, the complete rate-distortion region with arbitrary number of agents and SNR values was characterized by Prabhakaran\el \cite{Prabhakaran-2004-ISIT} and Oohama \cite{Oohama-2005-TIT}. More recently, Courtade and Weissman \cite{Courtade-2014-TIT} gave the rate-distortion region of the CEO problem under the logarithmic-loss distortion measure.

Secure lossless source coding with uncoded side information at the legitimate decoder and the eavesdropper was studied by Prabhakaran and Ramchandran \cite{Prabhakaran-2007-ITW} with the assumption of no rate constraint on the encoder-decoder link. The minimum leakage rate was derived and it was shown that due to the side information at the eavesdropper, the usual Slepian-Wolf scheme \cite{Slepian-1973-TIT} is not always optimal. Lossless source coding with coded side information at the decoder (the so-called one-helper problem) and no side information at the eavesdropper was studied by Tandon\el \cite{Tandon-2013-TIT} where the rate-equivocation region was characterized. This setup was extended by G\"{u}nd\"{u}z\el \cite{Gunduz-2008-ITW} with additional side information at the eavesdropper in which inner and outer bounds on the compression-equivocation rate region were derived that did not match in general. Secure distributed lossless compression of two correlated sources, in which both sources were to be estimated at the decoder, was considered by Luh and Kundur \cite{Luh-2007-GLOBECOM} without side information at the eavesdropper and by G\"{u}nd\"{u}z\el \cite{Gunduz-2008-ISIT} with side information at the eavesdropper. These models were generalized by Salimi\el \cite{Salimi-2010-IET} to the case where both the legitimate receiver and the eavesdropper have access to correlated side information and the eavesdropper can choose to intercept either links from the encoders to the decoder at each instant. In \cite{Salimi-2010-IET}, inner and outer bounds for the compression-equivocation region were provided which were proved to be tight for several special cases.

The extension to the lossy case was considered in \cite{Luh-2008-TIFS, Villard-2010-Allerton, Ekrem-2011-Allerton}, and more recently by Villard and Piantanida \cite{Villard-2013-TIT} in which inner and outer bounds on the rate-distortion-equivocation region were derived. The optimal characterization of the rate-distortion-equivocation region was first found in \cite{Villard-2010-Allerton} for the lossy case with uncoded side information. Later in \cite{Villard-2013-TIT}, the optimal characterization for the lossless case was also derived. A different setup was considered by Kittichokechai\el  \cite{Kittichokechai-2013-ISIT} in which the eavesdropper can only access the coded side information, and the complete region was characterized under the logarithmic-loss distortion \cite{Courtade-2014-TIT}. Chia and Kittichokechai \cite{Chia-2013-ISIT} studied the case when the encoder has access to the side information of the decoder. Tandon\el \cite{Tandon-2013a-TIT} considered a scenario with two legitimate receivers and investigated the privacy of side information at one receiver with respect to the other one. An alternative approach to provide secrecy in source coding problems is based on having a shared secret key between the transmitter and the legitimate receiver \cite{Yamamoto-1994-TIT,Yamamoto-1997-TIT,Schieler-2014-TIT}, although we do not exploit this approach in our work.


\subsection{Contributions}
Our setup in this paper has two main distinctions from the aforementioned scenarios; first, the destination (CEO) is interested in estimation of the original source rather than the agents' observations as in all prior works. Similarly, the secrecy constraints in our problem are on the equivocation of the eavesdropper with respect to the remote source, not to the observations of the agents. In fact, our setup is a generalization of the previous cases considered for lossy secure source coding problems. We extend our previous work \cite{Naghibi-2013-ISWCS} for the lossless variant of this problem to the lossy case and derive inner and outer bounds on the rate-distortion-equivocation region of the CEO problem with secrecy constraints. We also investigate the region in special cases where the bounds are tight and we show that for these special cases our results coincide with the previous results in the literature.

In addition, we consider the quadratic Gaussian CEO problem with secrecy constraints and provide the optimal characterization of the rate-distortion-equivocation region for the case when the eavesdropper has no side information and an achievable region for a more general setup with side information at the eavesdropper.


\subsection{Notations and Organization}
In this paper, %
we use capital letters to indicate a random variable, small letters to indicate realization of a random variable, calligraphic letters to denote a set, e.g., $\calX$, and $|\calX|$ to indicate the cardinality of the set. The notation $X^n$ denotes the sequence $\{X_1,\dots,X_n\}$. The notion $X-Y-Z$ shows that $X$, $Y$, and $Z$ form a Markov chain, i.e., $p(x,y,z)=p(x,y)p(z|y)$ or $p(x,y,z)=p(x|y)p(y,z)$. We define $\calI_M\coloneqq\{1,\dots,M\}$ for $M\in\Natural$, and $[x]^+\coloneqq\max\{0,x\}$ for $x\in\Real$. Finally, $\mathbbm{1}_{\Real_{>0}}(x):\calX\to\{0,1\}$ denotes the indicator function such that $\mathbbm{1}_{\Real_{>0}}(x)=1$ for $x\in\Real_{>0}$, and $\mathbbm{1}_{\Real_{>0}}(x)=0$ otherwise.

The rest of the paper is organized as follows: In Section~\ref{sec:setup}, we describe the problem along with some definitions. Main results for inner and outer bounds on the rate-distortion-equivocation region are presented in Section~\ref{sec:mainRes}. Then, we study some special cases of our results in Section~\ref{sec:specialCase} where the region is completely characterized. The rate-distortion-equivocation region for the quadratic Gaussian case is given in Section~\ref{sec:gaussian}. Finally, the paper is concluded in Section~\ref{sec:conclusion}.


\section{Problem Setting}
\label{sec:setup}

\begin{figure}[t]
 \centering
 \includegraphics[width=0.93\columnwidth]{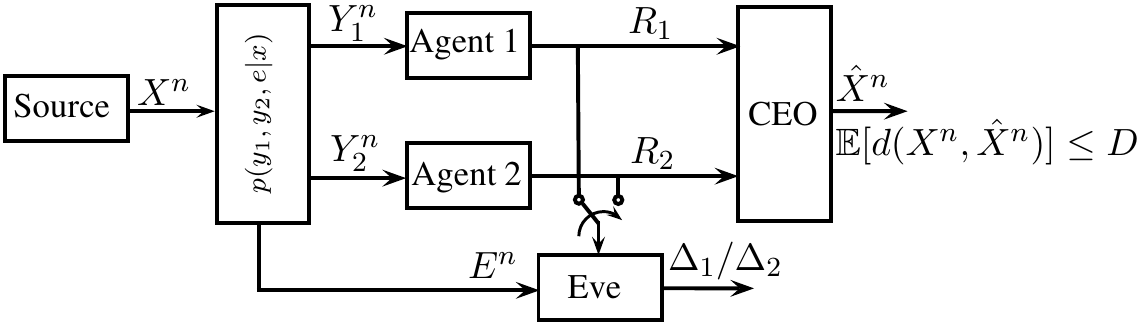}
 \caption{The CEO problem with secrecy constraints.}
 \label{fig:sysModel}
\end{figure}

We consider the CEO problem with secrecy constraints as depicted in \figref{fig:sysModel}. In this setup, two non-cooperative and independent agents have access to length-$n$ observations $Y_1^n$ and $Y_2^n$, respectively, which are noisy versions of the source sequence $X^n$. These observations are conditionally independent given $X^n$. Each agent independently transmits a compressed version of its observation to the CEO over a rate-limited noiseless link. The CEO estimates the source sequence $\hat{X}^n$ based on the received information from the two agents. An eavesdropper, referred to as Eve, with access to side information $E^n$ correlated to the source sequence $X^n$ can eavesdrop only one of the links from the agents to the CEO at each time instance to obtain as much information as possible about the source. Therefore,  agents' transmission rates should be such that the CEO can reconstruct the source reliably within a certain mean distortion threshold while simultaneously the equivocation at Eve is maximized. Eve's equivocation, with respect to either links, corresponds to her uncertainty about the original source when she combines her side information with the information obtained from the link. We assume that Eve cannot access both links simultaneously as the links are noise-free and in such case she would be more powerful than the CEO for estimating the source due to her additional side information. The sequences $X^n$, $Y_1^n$, $Y_2^n$, and $E^n$ are independent and identically distributed (i.i.d.) according to joint distribution $p(x,y_1,y_2,e)=p(x)p(y_1|x)p(y_2|x)p(e|x)$ over the finite alphabet $\calX \times \calY_1 \times \calY_2 \times \calE$.

Let $d\!:\!\calX\!\times\!\calX\!\rightarrow\![0,d_{\max}]$ be a finite distortion measure. %
We define the component-wise mean distortion between two sequences $x^n$, $\hat{x}^n$ in $\calX^n$ as 
\begin{equation}
d^{(n)}(x^n,\hat{x}^n)\coloneqq \frac{1}{n}\sum_{i=1}^n d(x_i,\hat{x}_i).
\end{equation}

\begin{definition}
\label{def:code}
A $(M_1,M_2,n)$-code for compression and transmission of the source by the agents with secrecy constraints  consists of an encoding function at each agent, $f_j:\calY_j^n\rightarrow\calI_{M_j}$ for $j\in\{1,2\}$, and a decoding function at the CEO, $g:\calI_{M_1}\times\calI_{M_2}\rightarrow\calX^n$. The equivocation rates for this code are defined as $\frac{1}{n}H\big(X^n|f_j(Y_j^n),E^n\big)$ for $j\in\{1,2\}$.
\end{definition}

\begin{definition}
\label{def:achR}
A tuple $(R_1,R_2,\Delta_1,\Delta_2,D)$ is said to be achievable if $\forall\epsilon>0$ there exists $N_0\in\Natural_{>0}$ such that for all $n>N_0$ there exists a sequence of $(M_1,M_2,n)$-codes with
\begin{align*}
\log(M_j)&\leq n(R_j+\epsilon),~ \forall j\in\{1,2\},  \\
H\big(X^n|f_j(Y_j^n),E^n\big)&\geq n(\Delta_j-\epsilon),~ \forall j\in\{1,2\}, \\
\E\Big[d^{(n)}\big(X^n,g(f_1(Y_1^n),f_2(Y_2^n))\big)\Big] &\leq D+\epsilon.
\end{align*}
Let $\calR$ denote the rate-distortion-equivocation region defined as the set of all achievable tuples $(R_1,R_2,\Delta_1,\Delta_2,D)$.
\end{definition}


\section{Inner and Outer Bounds on the Rate-Distortion-Equivocation Region}
\label{sec:mainRes}

\subsection{Inner Bound}

\begin{theorem}
\label{th:inner}
Let $\calR_{\text{in}}$ denote the region defined as the closure of the convex hull of the set of all tuples $(R_1,R_2,\Delta_1,\Delta_2,D)$ such that there exist random variables $V_1$, $V_2$, $U_1$, and $U_2$ on some finite sets $\calV_1$, $\calV_2$, $\calU_1$, and $\calU_2$, respectively, according to the distribution $p(x,y_1,y_2,e,v_1,v_2,u_1,u_2) = p(x)p(y_1|x)p(y_2|x) p(e|x) p(u_1|y_1)p(u_2|y_2)p(v_1|u_1)p(v_2|u_2)$ and a function $\hat{X}:\calU_1\times\calU_2\rightarrow \calX$ that satisfy
\begin{align}
R_1  &\geq  I(U_1;Y_1|U_2), \label{eq:R1_inner}\\
R_2  &\geq  I(U_2;Y_2|U_1),\\
R_1 + R_2  &\geq  I(U_1,U_2;Y_1,Y_2), \label{eq:Rsum_inner}\\
\Delta_1  &\leq  \big[H(X|V_1,E)- I(U_1;Y_1|V_1,U_2) \nonumber\\
&\qquad  + I(U_1;Y_1|V_1,X)\big]^+, \label{eq:delta1_inner}\\
\Delta_2  &\leq  \big[H(X|V_2,E) -  I(U_2;Y_2|V_2,U_1) \nonumber\\
&\qquad + I(U_2;Y_2|V_2,X)\big]^+, \label{eq:delta2_inner} \\ 
\Delta_1  +  \Delta_2  &\leq  \big[H(X|V_1,E) +  H(X|V_2,E) \nonumber\\
&\qquad - I(U_1,U_2;Y_1,Y_2|V_1,V_2) \nonumber\\
&\qquad + I(U_1;Y_1|V_1,X) + I(U_2;Y_2|V_2,X)\big]^+, \label{eq:delta12_inner} \\
\Delta_1  - R_2  &\leq  \big[H(X|V_1,E) -I(U_2;Y_2|U_1)  \nonumber\\
&\qquad  -  I(U_1;Y_1|V_1) + I(U_1;Y_1|V_1,X)\big]^+, \label{eq:delta1R2_inner} \\
\Delta_2 - R_1  &\leq  \big[H(X|V_2,E) - I(U_1;Y_1|U_2) \nonumber\\
&\qquad -  I(U_2;Y_2|V_2) + I(U_2;Y_2|V_2,X)\big]^+,  \label{eq:delta2R1_inner} \\
D&\geq\E\Big[d\big(X,\hat{X}(U_1,U_2)\big)\Big]. \label{eq:D_inner}
\end{align}
Then, we have $\calR_{\text{in}}\subset\calR$.
\end{theorem}

\begin{IEEEproof}
The proof is given in Appendix~\ref{sec:proof_inner}.
\end{IEEEproof}

\begin{proposition}
\label{prop:card_inner}
In Theorem~\ref{th:inner}, it suffices to consider auxiliary random variables $V_j$ and $U_j$ for $j\in\{1,2\}$ with cardinalities $|\calV_j|\leq |\calY_j|+9$ and $|\calU_j|\leq (|\calY_j|+9)(|\calY_j|+5)$, respectively (see Appendix~\ref{sec:proof_card_inner} for the proof).
\end{proposition}

The achievability scheme resulting in the inner bound is based on superposition coding and random binning at the agents, and joint decoding at the CEO. In particular, agent $j$ first transmits the bin index related to the auxiliary random variable $V_j$ with distribution $p(v_j|u_j)$ via the noiseless link. Then, the agents send the remaining information which is required for the CEO to be able to reconstruct the source based on the Wyner-Ziv scheme \cite{Wyner-1976-TIT}. The detailed proof is given in Appendix~\ref{sec:proof_inner}, however, we provide some intuitions on the results. %
Inequalities \eqref{eq:R1_inner}--\eqref{eq:Rsum_inner} and \eqref{eq:D_inner} are similar to the Berger-Tung bounds \cite{Berger_MSC,Tung_PhD} that establish perfect estimation of $U_1$ and $U_2$ at the CEO from which $X$ can be reconstructed within the distortion limit $D$. In the equivocation bounds \eqref{eq:delta1_inner} and \eqref{eq:delta2_inner}, the first term corresponds to Eve's uncertainty about the source after decoding the codeword $v_j^n$ based on the received bin index combined with her side information and the second term is the reduction in her uncertainty when receiving the remaining information transmitted to the CEO by the agents. Finally, the last term in \eqref{eq:delta1_inner} and \eqref{eq:delta2_inner} stems from the fact that in contrast to previous works, the secrecy constraints are on Eve's equivocation with respect to the original source while the transmitted information by the agents are functions of their respective observations and not the source, resulting in an increase in Eve's uncertainty. Inequalities \eqref{eq:delta1R2_inner} and \eqref{eq:delta2R1_inner} depict a trade-off between Eve's equivocation and transmission rates, implying that each link's transmission rate limits the other link's equivocation rate.

\begin{remark}
The region of Theorem~\ref{th:inner} can also be obtained by constructing six different codes achieving the corner points shown in Tables~\ref{tab:corners1}--\ref{tab:corners2} and using the time-sharing technique between these points. Each corner point is achieved using a four-step communication to transmit variables $V_1$, $V_2$, $U_1$, and $U_2$ to the CEO with different decoding orders, provided that $V_j$ is decoded prior to $U_j$ for $j\in\{1,2\}$. In each step, previously received and decoded information at the CEO is used as side information for the current decoding step. Each code employs superposition coding, with $V_j$ as the first layer and $U_j$ as the second layer, and random binning based on the available side information at the CEO in each communication step.
\end{remark}

\begin{table*}[t]
\renewcommand{\arraystretch}{1.3}
\centering
\caption{Corner points of the inner region corresponding to different decoding orders: rates and distortion.}
\label{tab:corners1}
\begin{tabular}{@{} c c c c c @{}} \toprule
Corner point & Decoding order & $R_1$ & $R_2$ & $D$ \\ \midrule
1 & $V_2,U_2,V_1,U_1$ & $I(U_1;Y_1|U_2)$ & $I(U_2;Y_2)$ & $\E\big[d\big(X,\hat{X}(U_1,U_2)\big)\big]$ \\
2 & $V_2,V_1,U_2,U_1$ & $I(V_1;Y_1|V_2)+I(U_1;Y_1|V_1,U_2)$ & $I(V_2;Y_2)+I(U_2;Y_2|V_1,V_2)$ & $\E\big[d\big(X,\hat{X}(U_1,U_2)\big)\big]$ \\
3 & $V_1,V_2,U_2,U_1$ & $I(V_1;Y_1)+I(U_1;Y_1|V_1,U_2)$ & $I(U_2;Y_2|V_1)$ & $\E\big[d\big(X,\hat{X}(U_1,U_2)\big)\big]$ \\
4 & $V_1,U_1,V_2,U_2$ & $I(U_1;Y_1)$ & $I(U_2;Y_2|U_1)$ & $\E\big[d\big(X,\hat{X}(U_1,U_2)\big)\big]$\\
5 & $V_1,V_2,U_1,U_2$ & $I(V_1;Y_1)+I(U_1;Y_1|V_1,V_2)$ & $I(V_2;Y_2|V_1)+I(U_2;Y_2|V_2,U_1)$ & $\E\big[d\big(X,\hat{X}(U_1,U_2)\big)\big]$\\
6 & $V_2,V_1,U_1,U_2$ & $I(U_1;Y_1|V_2)$ & $I(V_2;Y_2)+I(U_2;Y_2|V_2,U_1)$ & $\E\big[d\big(X,\hat{X}(U_1,U_2)\big)\big]$\\
\bottomrule
\end{tabular}
\end{table*}

\begin{table*}[t]
\renewcommand{\arraystretch}{1.3}
\centering
\caption{Corner points of the inner region corresponding to different decoding orders: equivocation rates.}
\label{tab:corners2}
\begin{tabular}{@{} c c c c @{}} \toprule
Corner point & Decoding order & $\Delta_1$ & $\Delta_2$ \\ \midrule
1 & $V_2,U_2,V_1,U_1$ & $\big[H(X|V_1,E)- I(U_1;Y_1|V_1,U_2) + I(U_1;Y_1|V_1,X)\big]^+$ & $\big[H(X|V_2,E) -  I(U_2;Y_2|V_2) + I(U_2;Y_2|V_2,X)\big]^+$ \\
2 & $V_2,V_1,U_2,U_1$ & $\big[H(X|V_1,E)- I(U_1;Y_1|V_1,U_2) + I(U_1;Y_1|V_1,X)\big]^+$ & $\big[H(X|V_2,E) -  I(U_2;Y_2|V_1,V_2) + I(U_2;Y_2|V_2,X)\big]^+$ \\
3 & $V_1,V_2,U_2,U_1$ & $\big[H(X|V_1,E)- I(U_1;Y_1|V_1,U_2) + I(U_1;Y_1|V_1,X)\big]^+$ & $\big[H(X|V_2,E) -  I(U_2;Y_2|V_1,V_2) + I(U_2;Y_2|V_2,X)\big]^+$ \\
4 & $V_1,U_1,V_2,U_2$ & $\big[H(X|V_1,E)- I(U_1;Y_1|V_1) + I(U_1;Y_1|V_1,X)\big]^+$ & $\big[H(X|V_2,E) -  I(U_2;Y_2|V_2,U_1) + I(U_2;Y_2|V_2,X)\big]^+$ \\
5 & $V_1,V_2,U_1,U_2$ & $\big[H(X|V_1,E)- I(U_1;Y_1|V_1,V_2) + I(U_1;Y_1|V_1,X)\big]^+$ & $\big[H(X|V_2,E) -  I(U_2;Y_2|V_2,U_1) + I(U_2;Y_2|V_2,X)\big]^+$ \\
6 & $V_2,V_1,U_1,U_2$ & $\big[H(X|V_1,E)- I(U_1;Y_1|V_1,V_2) + I(U_1;Y_1|V_1,X)\big]^+$ & $\big[H(X|V_2,E) -  I(U_2;Y_2|V_2,U_1) + I(U_2;Y_2|V_2,X)\big]^+$ \\
\bottomrule
\end{tabular}
\end{table*}


\subsection{Outer Bound}

\begin{theorem}
\label{th:outer}
Let $\calR_{\text{out}}$ denote the region defined as the closure of the set of all tuples $(R_1,R_2,\Delta_1,\Delta_2,D)$ such that there exist random variables $V_1$, $V_2$, $U_1$, and $U_2$ on some finite sets $\calV_1$, $\calV_2$, $\calU_1$, and $\calU_2$, respectively, which form Markov chains $V_j-U_j-Y_j-(X,E,Y_{j'})$ for $j,j'\in\{1,2\}$ with $j\neq j'$, and a function $\hat{X}:\calU_1\times\calU_2\rightarrow \calX$ that satisfy
\begin{align}
R_1  &\geq  I(U_1;Y_1|U_2), \label{eq:R1_outer}\\
R_2  &\geq  I(U_2;Y_2|U_1), \label{eq:R2_outer}\\
R_1 + R_2  &\geq  I(U_1,U_2;Y_1,Y_2), \label{eq:Rsum_outer}\\
\Delta_1  &\leq  H(X|E)- I(X;V_1|E), \label{eq:delta1_outer}\\
\Delta_2  &\leq  H(X|E)- I(X;V_2|E), \label{eq:delta2_outer} \\
\Delta_1 - R_2  &\leq H(X|E)- I(X;V_1|E) - I(X;V_2|V_1,E), \label{eq:delta1R2_outer}\\
\Delta_2 - R_1  &\leq H(X|E)- I(X;V_2|E) - I(X;V_1|V_2,E), \label{eq:delta2R1_outer}\\
D&\geq\E\Big[d\big(X,\hat{X}(U_1,U_2)\big)\Big]. \label{eq:D_outer}
\end{align}
Then, we have $\calR\subset\calR_{\text{out}}$.
\end{theorem}

\begin{IEEEproof}
The proof is given in Appendix~\ref{sec:proof_outer}.
\end{IEEEproof}

\begin{proposition}
\label{prop:card_outer}
In Theorem~\ref{th:outer}, it suffices to consider auxiliary random variables $V_j$ and $U_j$ for $j\in\{1,2\}$ with cardinalities $|\calV_j|\leq |\calY_j|+7$ and $|\calU_j|\leq (|\calY_j|+7)(|\calY_j|+3)$, respectively (see Appendix~\ref{sec:proof_card_outer} for the proof).
\end{proposition}


\section{Special Case: The One-Helper Problem with Secrecy Constraints}
\label{sec:specialCase}

If Agent~1 has access to the source sequence $X^n$, our setup reduces to the lossy source coding problem with a helper and an eavesdropper who can choose to listen in on either source-destination or helper-destination links.
\begin{corollary}
\label{cor:one-helper}
In the above setting, if we additionally assume the helper's link is perfectly secure, our results coincide with the results given by Villard and Piantanida \cite[Theorem~3]{Villard-2013-TIT}. The inner bound is obtained by setting  $Y_1=X$, $V_2=U_2$, and removing the constraints on $\Delta_2$ in Theorem~\ref{th:inner}, and the outer bound can be proved similar to the proof of Theorem~\ref{th:outer}.
\end{corollary}

\begin{corollary}
\label{cor:uncoded_SI}
In the described one-helper problem with secrecy constraints, if $R_2\geq H(Y_2)$, the helper's sequence can be reconstructed by the destination losslessly. Then, the rate-distortion-equivocation region is characterized by
\begin{align}
R_1  &\geq  I(X;U_1|Y_2),\\
\Delta_1  &\leq  \big[H(X|V_1,E)- I(U_1;Y_1|V_1,Y_2) \big]^+, \\
\Delta_2  &\leq  H(X|Y_2,E),  \\
D&\geq\E\Big[d\big(X,\hat{X}(U_1,Y_2)\big)\Big],
\end{align}
\end{corollary}
where the auxiliary random variables $V_1$ and $U_1$ satisfy the Markov chain $V_1-U_1-X-(E,Y_2)$.

The achievability proof follows from the proof of Theorem~\ref{th:inner} by setting $Y_1=X$ and $V_2=U_2=Y_2$. Inequalities \eqref{eq:delta12_inner}--\eqref{eq:delta2R1_inner} are inactive for this setup. The converse proof is given in \cite[Theorem~3]{Villard-2013-TIT} for the secure lossy source coding with uncoded side information. Note that if Eve intercepts the helper's link, it can also reconstruct the helper's sequence $Y_2^n$ losslessly.

\begin{corollary}
\label{cor:uncoded_SI_lossless}
For the lossless one-helper setting, i.e., $D=0$, if $R_2\geq H(Y_2)$, the rate-equivocation region writes as:
\begin{align}
R_1  &\geq  H(X|Y_2),\\
\Delta_1  &\leq  I(X;Y_2|V_1) - I(X;E|V_1), \\
\Delta_2  &\leq  H(X|Y_2E).
\end{align}
\end{corollary}
The achievability proof follows from the proof of Theorem~\ref{th:inner} by setting $U_1=Y_1=X$ and $V_2=U_2=Y_2$. %
The converse proof is similar to the proof given in \cite{Prabhakaran-2007-ITW}. 
\begin{corollary}
\label{cor:no_Eve_SI_lossless}
For the lossless one-helper setting, i.e., $D=0$, if the eavesdropper has no side information, the rate-equivocation region is characterized by
\begin{align}
R_1  &\geq  H(X|U_2),\\
R_2  &\geq  I(Y_2;U_2),\\
\Delta_1  &\leq  I(X;U_2),\\
\Delta_2  &\leq  H(X|U_2).
\end{align}
\end{corollary}
The achievability is a special case of Theorem~\ref{th:inner} and obtained by setting $V_1$ and $E$ to be constants, $U_1=Y_1=X$, and $V_2=U_2$. The proof of converse is given in Appendix~\ref{sec:proof_no_Eve_SI}.

\section{The Quadratic Gaussian Case}
\label{sec:gaussian}

In this section, we study the Gaussian CEO problem with secrecy constraints and quadratic distortion measure. 

Let $X$ be a Gaussian source, i.e., $X\sim\calN(0,\sigma_X^2)$. The observations at the agents are modeled as $Y_j=X + N_j$ for $j\in\{1,2\}$, with $N_j\sim\calN(0,\sigma_{N_j}^2)$, where Gaussian random variables $X$, $N_1$, and $N_2$ are mutually independent.

\begin{figure}[t]
 \centering
 \includegraphics[width=0.92\columnwidth]{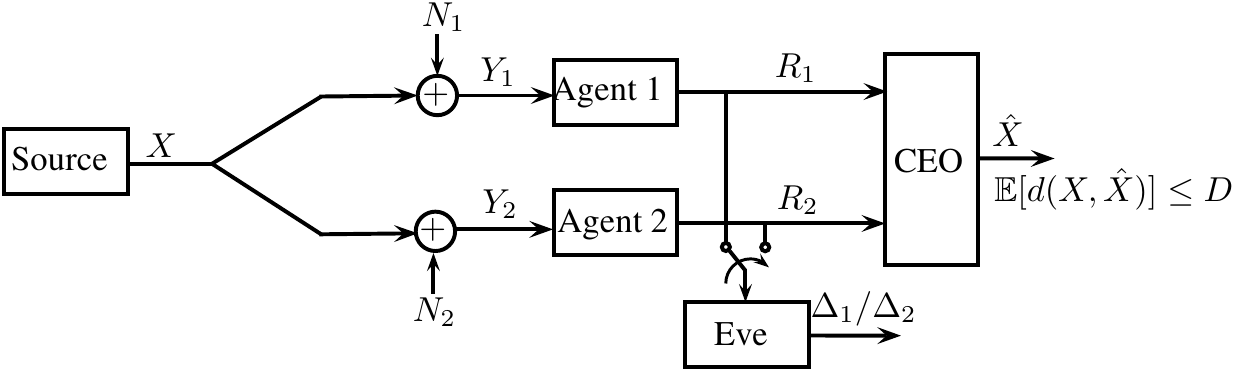}
 \vspace{-7pt}
 \caption{The quadratic Gaussian case with no side information at Eve.}
 \label{fig:sysModel_gauss_woSI}
 \vspace{-9pt}
\end{figure}

First, we consider the case where the eavesdropper has no side information. The model is depicted in \figref{fig:sysModel_gauss_woSI} and the following theorem provides the complete rate-distortion-equivocation region for this Gaussian setup.

\begin{theorem}
\label{th:gaussian}
In the quadratic Gaussian CEO problem with secrecy constraints, the rate-distortion-equivocation region is characterized by the set of all tuples $(R_1,R_2,\Delta_1,\Delta_2,D)$ satisfying
\begin{align}
R_1  &\geq  r_1 \!+\! \frac{1}{2}\log\frac{1}{D} \!-\! \frac{1}{2}\log \left( \frac{1}{\sigma_X^2}\!+\!\frac{1-2^{-2r_2}}{\sigma_{N_2}^2}\right), \label{eq:R1_gaussian_th}\\
R_2  &\geq  r_2 \!+\! \frac{1}{2}\log\frac{1}{D} \!-\! \frac{1}{2}\log \left( \frac{1}{\sigma_X^2}\!+\!\frac{1-2^{-2r_1}}{\sigma_{N_1}^2}\right), \label{eq:R2_gaussian_th}\\
R_1 + R_2  &\geq r_1 + r_2 + \frac{1}{2}\log\frac{1}{D} - \frac{1}{2}\log\frac{1}{\sigma_X^2}, \label{eq:R12_gaussian_th}\\
\Delta_1  &\leq \frac{1}{2}\log(2\pi e \sigma_X^2) - \frac{1}{2}\log\frac{1}{D}  \nonumber\\
&\qquad\qquad  + \frac{1}{2}\log \left( \frac{1}{\sigma_X^2}+\frac{1-2^{-2r_2}}{\sigma_{N_2}^2}\right), \label{eq:delta1_gaussian_th}\\
\Delta_2  &\leq \frac{1}{2}\log(2\pi e \sigma_X^2) - \frac{1}{2}\log\frac{1}{D} \nonumber\\
&\qquad\qquad + \frac{1}{2}\log \left( \frac{1}{\sigma_X^2}+\frac{1-2^{-2r_1}}{\sigma_{N_1}^2}\right),\label{eq:delta2_gaussian_th}\\
\Delta_1+\Delta_2  &\leq \log(2\pi e) - \frac{1}{2}\log\frac{1}{D} - \frac{1}{2}\log\frac{1}{\sigma_X^2},\label{eq:delta12_gaussian_th}\\
\Delta_1-R_2  &\leq \frac{1}{2}\log(2\pi e) - \frac{1}{2}\log\frac{1}{D} - r_2,\label{eq:delta1R2_gaussian_th}\\
\Delta_2-R_1  &\leq \frac{1}{2}\log(2\pi e) - \frac{1}{2}\log\frac{1}{D} - r_1,\label{eq:delta2R1_gaussian_th}
\end{align}
for some $(r_1,r_2)\in\Real^2_{\geq 0}$ that satisfy
\begin{align}
\frac{1}{D} &\leq \frac{1}{\sigma_X^2} + \frac{1-2^{-2r_1}}{\sigma_{N_1}^2} + \frac{1-2^{-2r_2}}{\sigma_{N_2}^2}. \label{eq:D_gaussian_th}
\end{align}
\end{theorem}

\begin{IEEEproof}
The proof is given in Appendix~\ref{sec:proof_gauss}.
\end{IEEEproof}

\begin{figure}[!t]
\centering
\begin{minipage}[t]{\columnwidth}
\centering
\subfloat{\includegraphics[width=0.48\columnwidth]{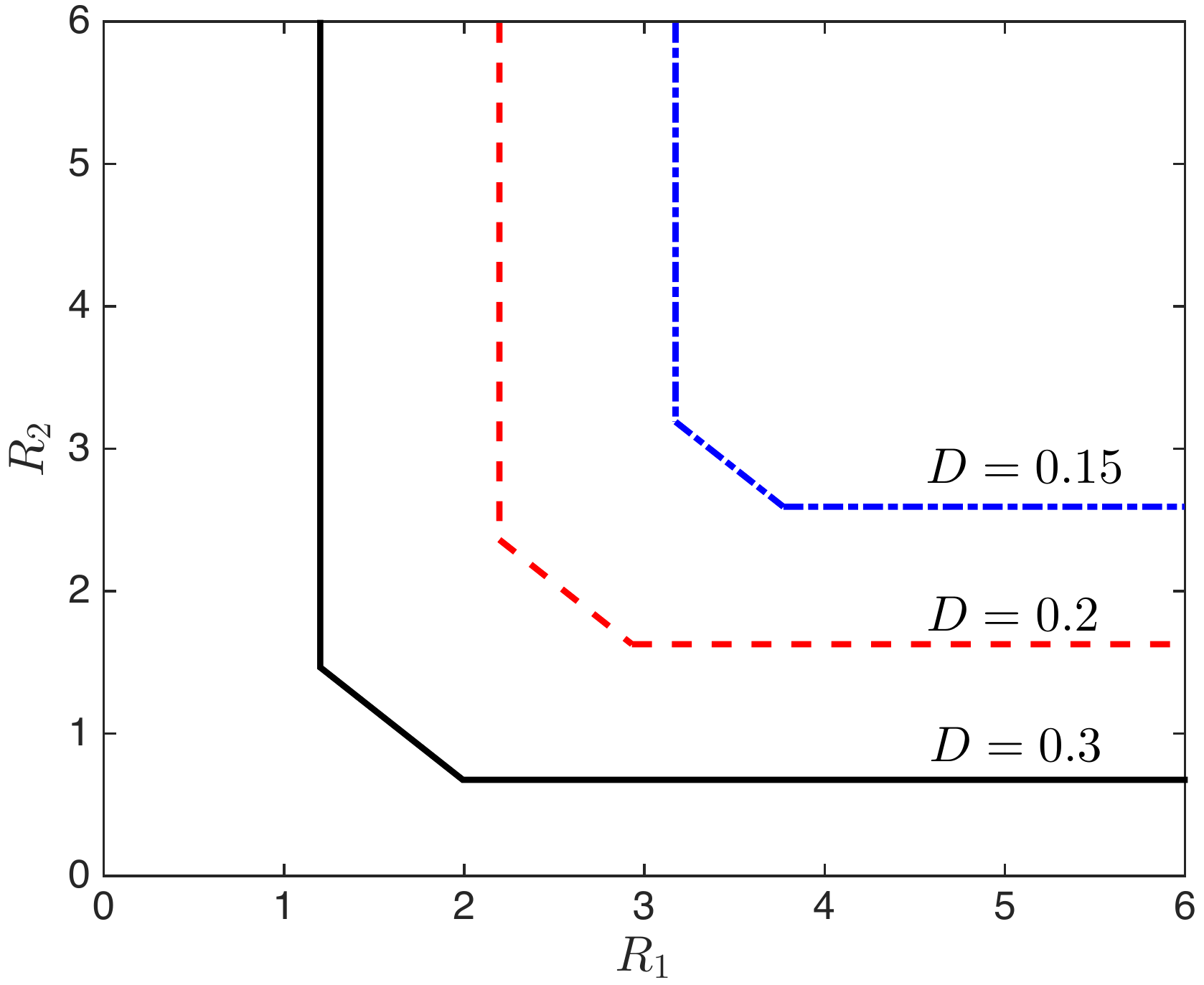}}
\hfill
\subfloat{\includegraphics[width=0.48\columnwidth]{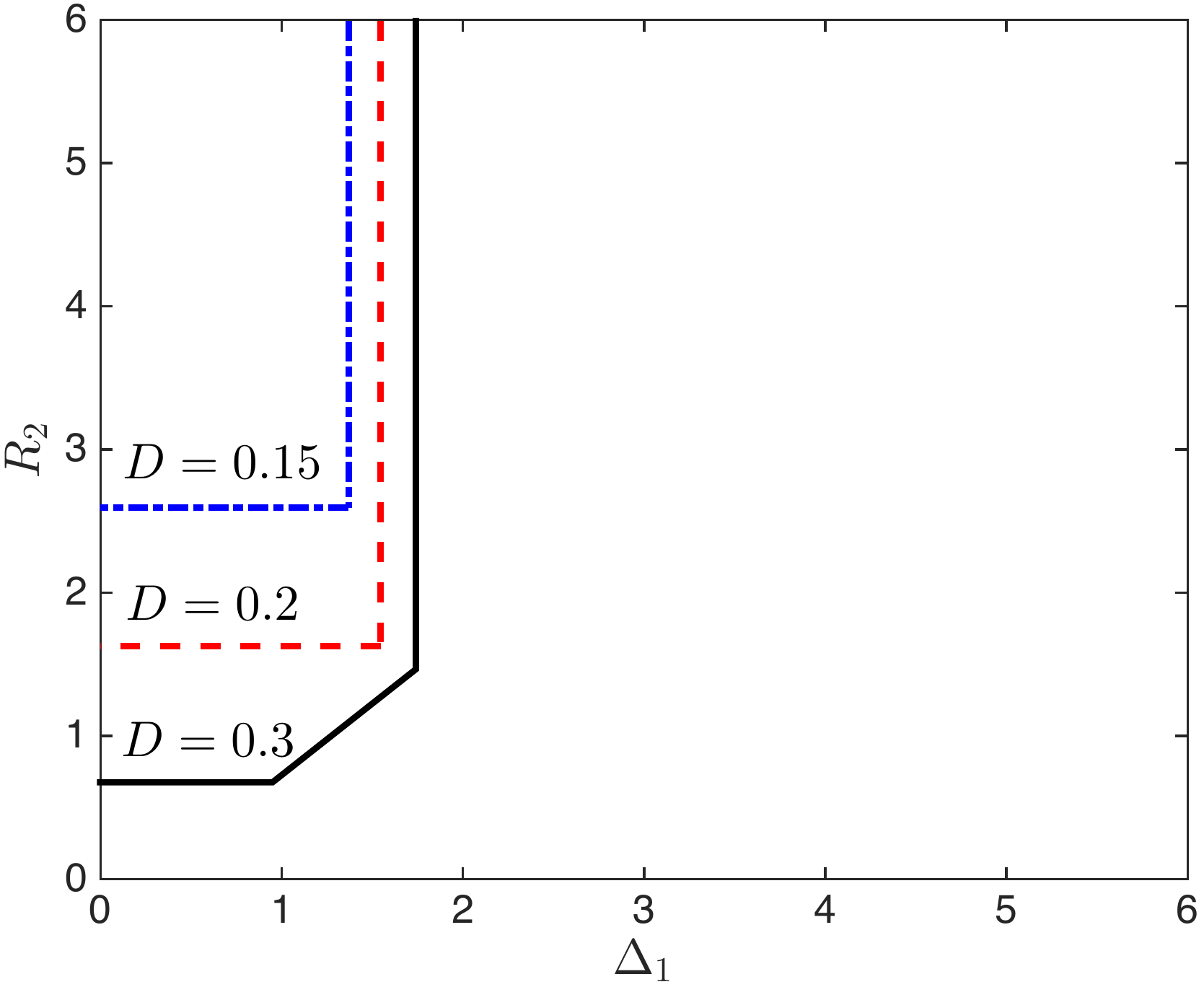}}
\end{minipage}\\
\vskip-5pt
\begin{minipage}[t]{\columnwidth}
\subfloat{\includegraphics[width=0.48\columnwidth]{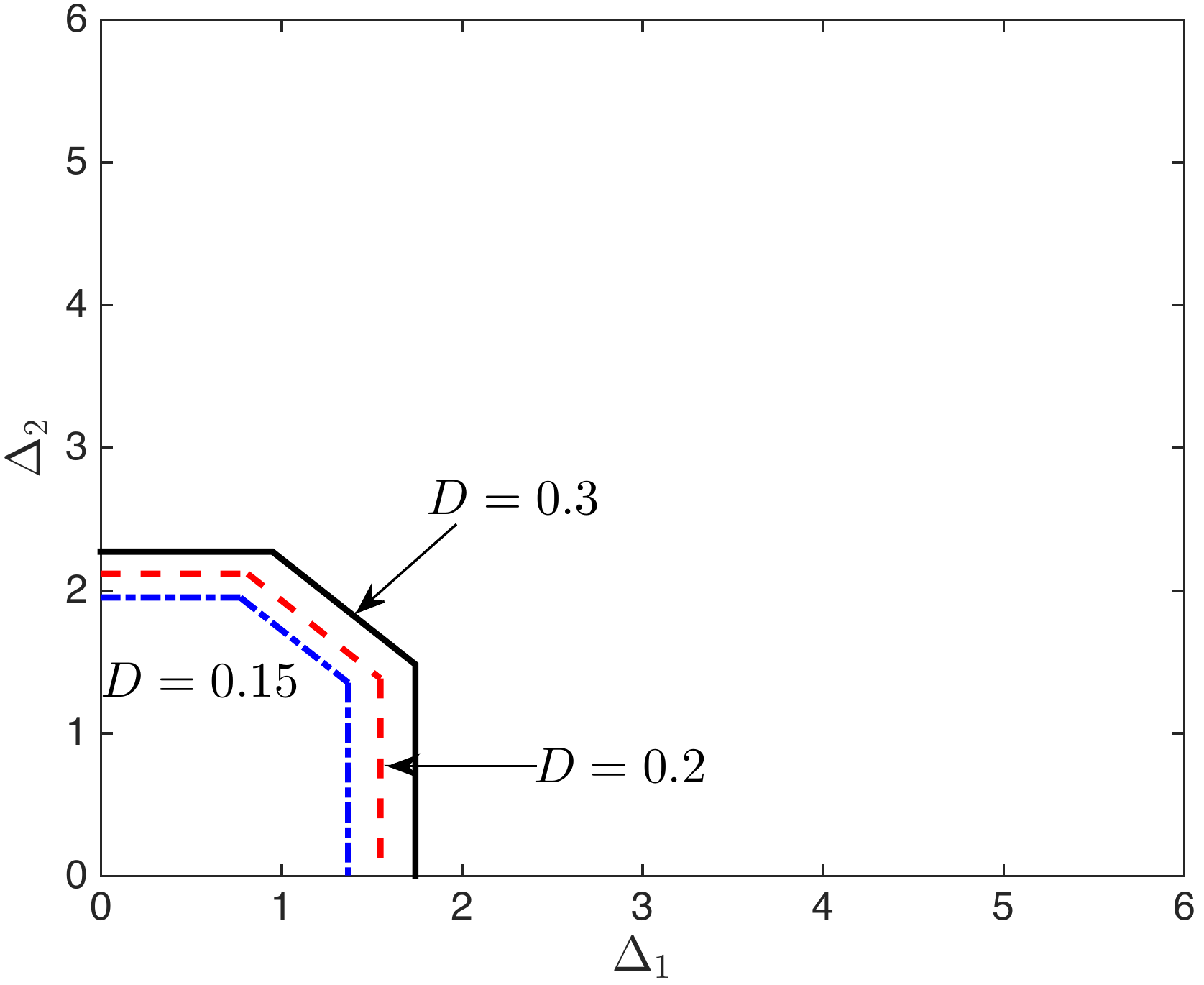}}
\hfill
\subfloat{\includegraphics[width=0.48\columnwidth]{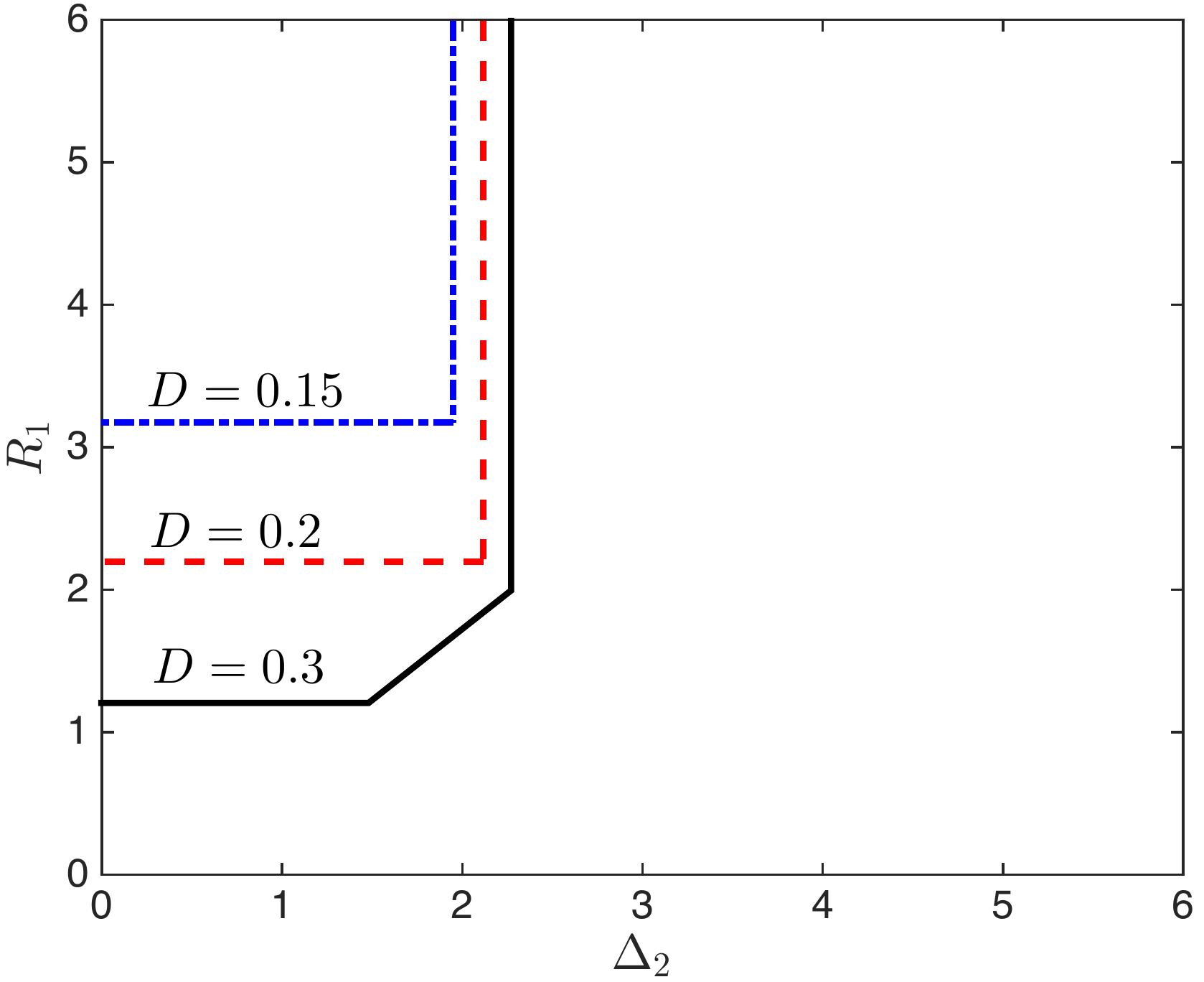}}
\vspace{-7pt}
\end{minipage}
\caption{An example of the rate-distortion-equivocation region for the quadratic Gaussian case with no side information at Eve and different distortion constraints.}
\label{fig_Gauss_region}
\vspace{-8pt}
\end{figure}

An example of the region of Theorem~\ref{th:gaussian} is illustrated in \figref{fig_Gauss_region} for different distortion constraints.

\begin{remark}
We note that equivocation as the secrecy measure in the finite alphabet setting represents the uncertainty of Eve about the source, but in the Gaussian setting, this interpretation is not quite valid. However, based on \cite[Theorem 8.6.6]{Cover_InformationTheory}, we can relate the equivocation rate (normalized differential entropy) to the estimation error at the eavesdropper. That is, we define the secrecy measure in the Gaussian setting as
\begin{align}
\frac{1}{n}\mathbb{E}\Big[\big\| X^n-\hat{Z}^{(n)}\big(f_j(Y_j^n),E^n\big)\big\|^2\Big] \geq \frac{1}{2\pi e}e^{2\Delta_j}\triangleq\theta_j,
\end{align}
for $j=1,2$, where $\hat{Z}^{(n)}\big(f_j(Y_j^n),E^n\big)$ is an estimator of $X^n$ at the eavesdropper. Then, we have  $\Delta_j=\frac{1}{2}\log(2\pi e \theta_j)$, and in this sense, equivocation rate provides a lower bound on the normalized distortion at Eve. Moreover, we can relate this to the information leakage rate as
\begin{align}
\frac{1}{n} I\big(X^n;f_j(Y_j^n,E^n)\big) &= h(X) - \frac{1}{n} h\big(f_j(Y_j^n,E^n)\big) \notag\\
&\leq h(X) - \Delta_j  \notag\\ 
&= h(X) - \frac{1}{2}\log(2\pi e \theta_j),
\end{align}
which is also in line with the result in \cite[Theorem~7.3]{Csiszar-2004-FTCIT}.
\end{remark}

\begin{figure}[t]
 \centering
 \includegraphics[width=0.95\columnwidth]{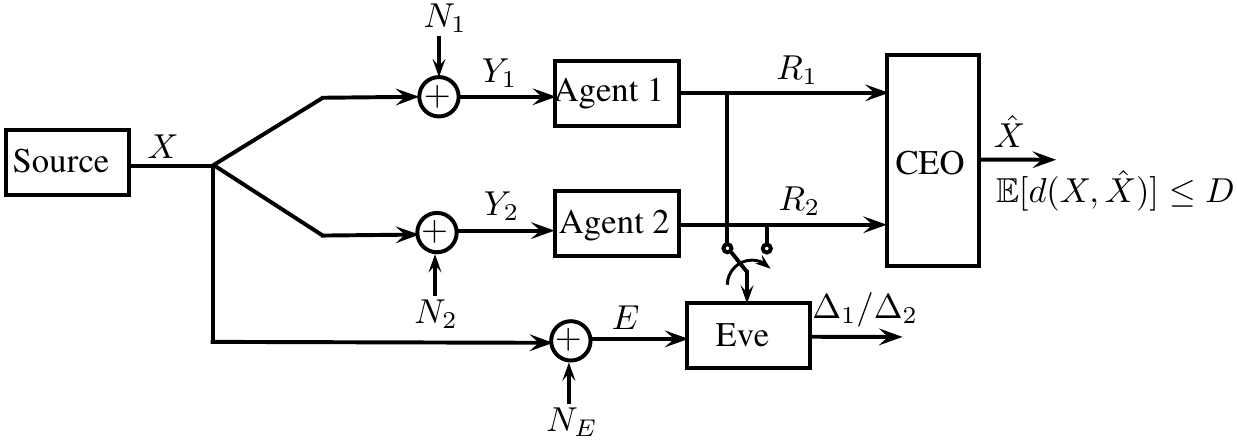}
 \caption{The quadratic Gaussian case with side information at Eve.}
 \label{fig:sysModel_gauss_wSI}
\end{figure}

Next, we consider the case where Eve has access to additional side information correlated to the source as shown in \figref{fig:sysModel_gauss_wSI}. %
We model this side information as $E=X+N_E$ where $N_E$ is a Gaussian random variable with $N_E\sim\calN(0,\sigma_{N_E}^2)$ and is independent of $X$, $N_1$, and $N_2$. The following theorem gives an inner bound for the rate-distortion-equivocation region of the quadratic Gaussian CEO problem with secrecy constraints and side information at the eavesdropper.

\begin{theorem}
\label{th:gaussian_SI}
In the quadratic Gaussian CEO problem with secrecy constraints and side information at the eavesdropper, a tuple $(R_1,R_2,\Delta_1,\Delta_2,D)$ is achievable if
\begin{align}
R_1  &\geq  r_1 \!+\! \frac{1}{2}\log\frac{1}{D} \!-\! \frac{1}{2}\log \left( \frac{1}{\sigma_X^2}\!+\!\frac{1-2^{-2r_2}}{\sigma_{N_2}^2}\right), \\
R_2  &\geq  r_2 \!+\! \frac{1}{2}\log\frac{1}{D} \!-\! \frac{1}{2}\log \left( \frac{1}{\sigma_X^2}\!+\!\frac{1-2^{-2r_1}}{\sigma_{N_1}^2}\right), \\
R_1 + R_2  &\geq r_1 + r_2 + \frac{1}{2}\log\frac{1}{D} - \frac{1}{2}\log\frac{1}{\sigma_X^2}, \\
\Delta_1  &\leq \frac{1}{2}\log(2\pi e) - \frac{1}{2}\log \left( \frac{1}{\sigma_X^2}\!+\!\frac{1}{\sigma_{N_E}^2}\right)  \nonumber\\
&\;  \!- \frac{1}{2}\log\frac{1}{D}\! +\! \frac{1}{2}\log \left( \frac{1}{\sigma_X^2}\!+\!\frac{1-2^{-2r_2}}{\sigma_{N_2}^2}\right) \! +\! T_1, \\
\Delta_2  &\leq \frac{1}{2}\log(2\pi e) - \frac{1}{2}\log \left( \frac{1}{\sigma_X^2}\!+\!\frac{1}{\sigma_{N_E}^2}\right)   \nonumber\\
&\; \!- \frac{1}{2}\log\frac{1}{D}\! +\! \frac{1}{2}\log \left( \frac{1}{\sigma_X^2}\!+\!\frac{1-2^{-2r_1}}{\sigma_{N_1}^2}\right)\! +\! T_2, \\
\Delta_1+\Delta_2  &\leq \log(2\pi e) - \log \left( \frac{1}{\sigma_X^2}\!+\!\frac{1}{\sigma_{N_E}^2}\right) - \frac{1}{2}\log\frac{1}{D}  \nonumber\\
&\; + \frac{1}{2}\log\frac{1}{\sigma_X^2} + T_1 + T_2 \nonumber\\
&\;  +  \frac{1}{2}\log \Bigg( \frac{\sigma_X^2\! +\! \frac{\sigma_{N_1}^2}{1-2^{-2r_1}}}{\frac{\sigma_{N_1}^2}{1-2^{-2r_1}}\!+\!\frac{\sigma_{N_2}^2}{1-2^{-2r_2}}} \Bigg)\mathbbm{1}_{\Real_{>0}}(T_1 \!+\! T_2),\\
\Delta_1-R_2  &\leq \frac{1}{2}\log(2\pi e) - \frac{1}{2}\log \left( \frac{1}{\sigma_X^2}\!+\!\frac{1}{\sigma_{N_E}^2}\right)  \nonumber\\
&\; - \frac{1}{2}\log\frac{1}{D} - r_2 + T_1 \nonumber\\
&\;  + \frac{1}{2}\log \Bigg( \frac{\sigma_X^2\! +\! \frac{\sigma_{N_1}^2}{1-2^{-2r_1}}}{\frac{\sigma_{N_1}^2}{1-2^{-2r_1}}\!+\!\frac{\sigma_{N_2}^2}{1-2^{-2r_2}}} \Bigg)\mathbbm{1}_{\Real_{>0}}(T_1),\\
\Delta_2-R_1  &\leq \frac{1}{2}\log(2\pi e) - \frac{1}{2}\log \left( \frac{1}{\sigma_X^2}\!+\!\frac{1}{\sigma_{N_E}^2}\right) \nonumber\\
&\; - \frac{1}{2}\log\frac{1}{D} - r_1 + T_2 \nonumber\\
&\;  +  \frac{1}{2}\log \Bigg( \frac{\sigma_X^2\! +\! \frac{\sigma_{N_1}^2}{1-2^{-2r_1}}}{\frac{\sigma_{N_1}^2}{1-2^{-2r_1}}\!+\!\frac{\sigma_{N_2}^2}{1-2^{-2r_2}}} \Bigg)\mathbbm{1}_{\Real_{>0}}(T_2),
\end{align}
where $\mathbbm{1}_{\Real_{>0}}(\cdot)$ is the indicator function and
\begin{align}
T_1 &\coloneqq \max\Bigg\{0, \frac{1}{2}\log \Bigg(1 + \frac{\frac{\sigma_{N_2}^2}{1-2^{-2r_2}}-\sigma_{N_E}^2}{\frac{\sigma_{N_1}^2}{1-2^{-2r_1}}+\sigma_{N_E}^2} \Bigg) \Bigg\}, \\
T_2 &\coloneqq \max\Bigg\{0, \frac{1}{2}\log \Bigg(1 + \frac{\frac{\sigma_{N_1}^2}{1-2^{-2r_1}}-\sigma_{N_E}^2}{\frac{\sigma_{N_2}^2}{1-2^{-2r_2}}+\sigma_{N_E}^2} \Bigg) \Bigg\},
\end{align}
for some $(r_1,r_2)\in\Real^2_{\geq 0}$ that satisfy
\begin{align}
\frac{1}{D} &\leq \frac{1}{\sigma_X^2} + \frac{1-2^{-2r_1}}{\sigma_{N_1}^2} + \frac{1-2^{-2r_2}}{\sigma_{N_2}^2}.
\end{align}
\end{theorem}

\begin{IEEEproof}
The proof is given in Appendix~\ref{sec:proof_gauss_wSI}.
\end{IEEEproof}

Note that if there is no correlation between Eve's side information and the source, i.e., $\sigma_{N_E}^2\to\infty$, the region of Theorem~\ref{th:gaussian_SI} coincides with the one in Theorem~\ref{th:gaussian}.

\section{Conclusion}
\label{sec:conclusion}
We studied the extension of the CEO problem with secrecy constraints. This setup is of interest to communication scenarios such as sensor networks or smart power grids in which links are vulnerable to eavesdropping. We derived inner and outer bounds on the rate-distortion-equivocation region in the discrete case. We also showed that the results that were derived for the one-helper problem with secrecy constraints in \cite{Tandon-2013-TIT} and \cite{Villard-2013-TIT} can be obtained as special cases of our results for the CEO problem with secrecy constraints. In addition, we provide the optimal region for the quadratic Gaussian case when Eve has no side information as well as an achievable region for a more general case. In this work, we have considered noise-free links from the agents to the CEO, however, it would be interesting to  investigate the effects of noisy channels in this problem. Moreover, extending this problem to include more agents and eavesdroppers with possibly different side information is another direction worthwhile investigating.

\appendices

\section{Proof of Theorem~\ref{th:inner}: The Inner Bound}
\label{sec:proof_inner}

We first state the following lemma that we use in the proof of Theorem~\ref{th:inner}. The lemma follows from \cite[Section~2.3]{Liang-2008-FTCIT}. 
\begin{lemma}
\label{lem:inner}
Let $\calC_n$ be a random codebook and $s$ be the corresponding codeword index of $V^n$ from this codebook. Let  $\mathrm{Pr}\left((V^n(s),E^n)\in\calT_\epsilon^{(n)}\right)\to 1$ as $n\to\infty$, where $\calT_\epsilon^{(n)}$ denotes the set of jointly $\epsilon$-typical $n$-sequences. Then, we have 
\begin{equation}
H(E^n|s,\calC_n)\leq n\big(H(E|V)+\epsilon\big). \label{eq:lem1}
\end{equation}
\end{lemma}
\begin{IEEEproof}[Proof of Lemma~\ref{lem:inner}]
Let $Z$ be a binary variable such that
\begin{equation*}
Z = \left\{\begin{array}{ll}
1 & \text{if } (V^n,E^n)\in \calT_\epsilon^{(n)}\\
0 & \text{otherwise.}\end{array}\right.
\end{equation*}
If $\mathrm{Pr}(Z=0)\leq\epsilon$ for sufficiently large $n$, we have
\begin{align}
H(E^n|s,\calC_n) &\leq H(E^n|V^n,\calC_n)\notag\\
&\leq H(E^n|V^n,Z) + H(Z)\notag\\
&\leq \mathrm{Pr}(Z=1)H(E^n|V^n,Z=1) \notag\\
&\qquad + \mathrm{Pr}(Z=0)H(E^n|V^n,Z=0) + H(Z)\notag\\
&\leq H(E^n|V^n,Z=1) + n\epsilon\log|\calE| + H(Z)\notag\\
&\leq \sum_{v^n\in\calT_\epsilon^{(n)}} p(v^n|Z=1)H(E^n|V^n=v^n,Z=1) \notag\\
&\qquad + n\epsilon\log|\calE| + H(Z)\notag\\
&\leq \sum_{v^n\in\calT_\epsilon^{(n)}} p(v^n|Z=1)\log|\calT_\epsilon^{(n)}(E|v^n)| \notag\\
&\qquad + n\epsilon\log|\calE| + H_2(\epsilon)\notag\\
&\leq n\big(H(E|V)+\epsilon'\big),
\end{align}
where $H_2(\cdot)$ is the binary entropy function. The last inequality is due to properties of jointly typical sequences \cite[Chapter~2]{ElGamal_NetInformationTheory}.
\end{IEEEproof}
Now, we proceed to prove Theorem~\ref{th:inner}.

Let $V_1$, $V_2$, $U_1$, and $U_2$  be random variables on some finite sets $\calV_1$, $\calV_2$, $\calU_1$, and $\calU_2$ according to the joint distribution $p(x,y_1,y_2,e,v_1,v_2,u_1,u_2) = p(x)p(y_1|x)p(y_2|x)p(e|x)$ $p(u_1|y_1)p(u_2|y_2)p(v_1|u_1)p(v_2|u_2)$, along with a function $\hat{X}:\calU_1\times\calU_2\rightarrow \calX$ satisfying the conditions of Theorem~\ref{th:inner}.

\emph{\textbf{Codebook generation}}: %
For fixed conditional distributions $p(u_j|y_j)$ and $p(v_j|u_j)$, $j=1,2$, randomly generate $2^{n(I(V_j;Y_j)+\epsilon_1)}$ independent codewords $v_j^n(s_j)$ of length $n$ according to $\prod_{i=1}^nP_{V_j}(v_{j,i}(s_j))$, where $s_j\in\{1,\dots,2^{n(I(V_j;Y_j)+\epsilon_1)}\}$. Then, divide them into $2^{n(R_{V_j}+\epsilon_2)}$ equal-sized bins, indexed by $b_j\in\{1,\dots,2^{n(R_{V_j}+\epsilon_2)}\}$ and denoted by $\{\calB_j(b_j)\}$. For each codeword $v_j^n(s_j)$, randomly generate $2^{n(I(U_j;Y_j|V_j)+\epsilon_3)}$ independent sequences $u_j^n(s_j,s'_j)$ according to $\prod_{i=1}^nP_{U_j}(u_{j,i}(s_j,s'_j))$, and divide them into $2^{n(R_{U_j}+\epsilon_4)}$  equal-sized bins, indexed by $w_j\in\{1,\dots,2^{n(R_{U_j}+\epsilon_4)}\}$ and denoted by $\{\calB'_j(s_j,w_j)\}$. Define $R_j= R_{V_j}+R_{U_j}$ for $j=1,2$. The codebook is revealed to the agents, CEO, and Eve.

\emph{\textbf{Encoding}}: %
Assume that the sequence $y_j^n$ is observed by Agent~$j$, $j=1,2$. Find a codeword $v_j^n(s_j)$ jointly typical with $y_j^n$. If there is more than one such codeword, select one  uniformly at random. If there is no such $v_j^n$, select one out of $2^{n(I(V_j;Y_j)+\epsilon_1)}$ uniformly at random. Given $v_j^n(s_j)$, find a codeword $u_j^n(s_j,s'_j)$ jointly typical with $y_j^n$. If there is more than one such codeword, select one  uniformly at random. If there is no such $u_j^n$, select one out of $2^{n(I(U_j;Y_j|V_j)+\epsilon_3)}$ uniformly at random.  The agent transmits the bin indices $b_j$ and $w_j$ of the codewords $v_j^n(s_j)\in\calB_j(b_j)$ and $u_j^n(s_j,s'_j)\in\calB'_j(s_j,w_j)$, respectively, i.e., $f_j(y_j^n)=(b_j,w_j)$. 

\emph{\textbf{Decoding at the CEO}}: %
Given the received messages from both agents, $J_1=(b_1,w_1)$ and $J_2=(b_2,w_2)$, find a unique index tuple $(\hat{s}_1,\hat{s}'_1,\hat{s}_2,\hat{s}'_2)$ such that the codewords $(v_1^n(\hat{s}_1),u_1^n(\hat{s}_1,\hat{s}'_1),v_2^n(\hat{s}_2),u_2^n(\hat{s}_2,\hat{s}'_2))$ are jointly typical, and they are in the bin indexed by $(b_1,w_1,b_2,w_2)$. %
If there is such a unique index tuple, compute the source estimate component-wise as $\hat{x}_i=g_i(J_1,J_2)\coloneqq\hat{X}\big(u_{1,i}(\hat{s}_1,\hat{s}'_1),u_{2,i}(\hat{s}_2,\hat{s}'_2)\big)$ for $i=1,\dots,n$; otherwise set the output to an arbitrary sequence in $\calX^n$. %

\emph{\textbf{Error analysis}}: %
Let $(s_1,s'_1,s_2,s'_2)$ and $(\hat{s}_1,\hat{s}'_1,\hat{s}_2,\hat{s}'_2)$ be the chosen indices at the encoders and decoder, respectively. Let $\mathrm{Pr}(\mathsf{E})$ denote the probability of an error event during encoding and decoding steps. %
We now show that this probability, averaged over all possible codebooks, tends to zero as $n\to\infty$ provided that conditions of Theorem~\ref{th:inner} is satisfied. Consider the following error events in the encoding steps (for $j,j'=1,2$, and $j\neq j'$):
\begin{align*}
&\mathsf{E}_0 \!=\! \Big\{\big(S^n,Y_1^n,Y_2^n,E^n\big)\notin \calT_\delta^{(n)}(S,Y_1,Y_2,E)\Big\},\\
&\mathsf{E}_1 \!=\! \mathsf{E}_0^c \cap \Big\{\big(V_j^n(\tilde{s}_j),Y_j^n\big)\notin \calT_\delta^{(n)}(V_j,Y_j)\Big\},\\ 
&\qquad\qquad\qquad\qquad \forall \tilde{s}_j\in\{1,\dots,2^{n(I(V_j;Y_j)+\epsilon_1)}\},\\
&\mathsf{E}_2 \!=\! \mathsf{E}_0^c \cap \mathsf{E}_1^c \cap \Big\{\big(U_j^n(s_j,\tilde{s}'_j),Y_j^n\big) \notin \calT_\delta^{(n)}(U_j,Y_j|v_j^n(s_j))\Big\},\\ 
&\qquad\qquad\qquad\qquad \forall \tilde{s}'_j\in\{1,\dots,2^{n(I(U_j;Y_j|V_j)+\epsilon_3)}\},\\
&\mathsf{E}_3 \!=\! \Big(\bigcap_{t=0}^2 \mathsf{E}_t^c \Big) \!\cap\! \Big\{\big(V_j^n(s_j),Y_1^n,Y_2^n\big)\notin \calT_\delta^{(n)}(V_j,Y_1,Y_2)\Big\},\\
&\mathsf{E}_4 \!=\! \Big(\bigcap_{t=0}^3 \mathsf{E}_t^c \Big) \!\cap\! \Big\{\big(U_j^n(s_j,s'_j),Y_1^n,Y_2^n\big) \\ 
&\qquad\qquad\qquad\qquad  \notin \calT_\delta^{(n)}(U_j,Y_1,Y_2|v_j^n(s_j))\Big\},\\
&\mathsf{E}_5 \!=\! \Big(\bigcap_{t=0}^4 \mathsf{E}_t^c \Big) \!\cap\! \Big\{\big(V_j^n(s_j),U_{j'}^n(s_{j'},s'_{j'}),Y_1^n,Y_2^n\big) \\ 
&\qquad\qquad\qquad\qquad \notin \calT_\delta^{(n)}(V_j,U_{j'},Y_1,Y_2|v_{j'}^n(s_{j'}))\Big\},\\
&\mathsf{E}_6 \!=\! \Big(\bigcap_{t=0}^5 \mathsf{E}_t^c \Big) \!\cap\! \Big\{\big(V_1^n(s_1),V_2^n(s_2),Y_1^n,Y_2^n\big) \\ 
&\qquad\qquad\qquad\qquad  \notin \calT_\delta^{(n)}(V_1,V_2,Y_1,Y_2)\Big\},\\
&\mathsf{E}_7 \!=\! \Big(\bigcap_{t=0}^6 \mathsf{E}_t^c \Big) \!\cap\! \Big\{\big(U_1^n(s_1,s'_1),U_2^n(s_2,s'_2),Y_1^n,Y_2^n\big) \\ 
&\qquad\qquad\qquad\qquad  \notin \calT_\delta^{(n)}(U_1,U_2,Y_1,Y_2|v_1^n(s_1),v_2^n(s_2))\Big\}.
\end{align*}
Next, consider the following error event in decoding step:
\begin{align*}
&\mathsf{E}_8 \!=\! \Big(\bigcap_{t=0}^7 \mathsf{E}_t^c \Big) \!\cap\! \Big\{\big(V_1^n(\tilde{s}_1),U_1^n(\tilde{s}_1,\tilde{s}'_1),V_2^n(\tilde{s}_2),U_2^n(\tilde{s}_2,\tilde{s}'_2)\big) \\ 
&\qquad\qquad  \in \calT_\delta^{(n)}(V_1,U_1,V_2,U_2), \\ 
&\qquad\qquad  \text{for some } (\tilde{s}_1,\tilde{s}'_1,\tilde{s}_2,\tilde{s}'_2)\neq (s_1,s'_1,s_2,s'_2), \\ 
&\qquad\qquad  \text{ s.t. } \big(V_1^n(\tilde{s}_1),U_1^n(\tilde{s}_1,\tilde{s}'_1),V_2^n(\tilde{s}_2),U_2^n(\tilde{s}_2,\tilde{s}'_2)\big) \\ 
&\qquad\qquad \in \calB_1(b_1)\times\calB'_1(\tilde{s}_1,w_1)\times\calB_2(b_2)\times\calB'_2(\tilde{s}_2,w_2) \Big\}.
\end{align*}
Finally, by the union of events bound, the probability of error in the encoding and decoding steps is upper bounded as
\begin{equation}
\mathrm{Pr}(\mathsf{E})\leq\mathrm{Pr}\Big(\bigcup_{t=0}^8 \mathsf{E}_t\Big) \leq \sum_{t=0}^8\mathrm{Pr}(\mathsf{E}_t). \label{eq:union_bound}
\end{equation}

We proceed to bound each term in \eqref{eq:union_bound}. From properties of typical sequences, $\mathrm{Pr}(\mathsf{E}_0)$ vanishes as $n\to\infty$. By covering lemma \cite[Lemma~3.3]{ElGamal_NetInformationTheory}, $\mathrm{Pr}(\mathsf{E}_1)$ and $\mathrm{Pr}(\mathsf{E}_2)$ tend to zero as $n\to\infty$. %
For $j,j'=1,2$, and $j\neq j'$, since $\big\{Y_{j'}^n|V_j^n(s_j)=v_j^n,Y_j^n=y_j^n\big\}\sim\prod_{i=1}^np(y_{j',i}|y_{j,i})$, by conditional typicality lemma \cite[Chapter~2]{ElGamal_NetInformationTheory}, $\mathrm{Pr}(\mathsf{E}_3)$ tends to zero as $n\to\infty$. Similarly, as $\big\{Y_{j'}^n|V_j^n(s_j)=v_j^n,U_j^n(s_j,s'_j)=u_j^n,Y_j^n=y_j^n\big\}\sim\prod_{i=1}^np(y_{j',i}|y_{j,i})$, $\mathrm{Pr}(\mathsf{E}_4)$ also vanishes as $n\to\infty$. %
To bound $\mathrm{Pr}(\mathsf{E}_5)$, let $(v_{j'}^n,u_{j'}^n,y_1^n,y_2^n)\in\calT_\delta^{(n)}(V_{j'},U_{j'},Y_1,Y_2)$. Then, $\mathrm{Pr}\big(V_j^n(s_j)|V_{j'}^n(s_{j'})=v_{j'}^n,U_j^n(s_j,s'_j)=u_{j'}^n,Y_1^n=y_1^n,Y_2^n=y_2^n)=p(v_j^n|y_j^n\big)$, and by Markov lemma \cite[Lemma~12.1]{ElGamal_NetInformationTheory}, $\mathrm{Pr}(\mathsf{E}_5)$ tends to zero as $n\to\infty$. Using similar steps and based on Markov lemma, $\mathrm{Pr}(\mathsf{E}_6)$ and $\mathrm{Pr}(\mathsf{E}_7)$ also tend to zeros as $n\to\infty$. %

As can be seen from $\mathsf{E}_8$, in the decoding step, an error occurs if the decoded codewords are jointly typical and they are in the bin indexed by $(b_1,w_1,b_2,w_2)$, however, the decoded tuple $(\hat{s}_1,\hat{s}'_1,\hat{s}_2,\hat{s}'_2)$ of codeword indices are different from the chosen ones at the encoders, i.e., $(s_1,s'_1,s_2,s'_2)$. We split this event into eight possible events (other events result in the same constraints as one of these eight events) and bound its probability using the union of events bound as follows:
\begin{align}
\mathrm{Pr}(\mathsf{E}_8)&=\mathrm{Pr}\big((\hat{s}_1,\hat{s}'_1,\hat{s}_2,\hat{s}'_2)\neq (s_1,s'_1,s_2,s'_2)\big)\label{eq:pd}\\
&\leq \mathrm{Pr}(\hat{s}_1\neq s_1) \label{eq:pd1}\\
&\;\;\; + \mathrm{Pr}(\hat{s}_1=s_1,\hat{s}_2 \neq s_2) \label{eq:pd2}\\
&\;\;\; + \mathrm{Pr}(\hat{s}_1 \neq s_1,\hat{s}_2=s_2) \label{eq:pd3}\\
&\;\;\; + \mathrm{Pr}(\hat{s}_1=s_1,\hat{s}'_1\neq s'_1,\hat{s}_2 = s_2) \label{eq:pd4}\\
&\;\;\; + \mathrm{Pr}(\hat{s}_1=s_1,\hat{s}'_1= s'_1,\hat{s}_2 \neq s_2) \label{eq:pd5}\\
&\;\;\; + \mathrm{Pr}(\hat{s}_1\neq s_1,\hat{s}_2 = s_2,\hat{s}'_2= s'_2) \label{eq:pd6}\\
&\;\;\; + \mathrm{Pr}(\hat{s}_1= s_1,\hat{s}'_1= s'_1,\hat{s}_2 = s_2,\hat{s}'_2\neq s'_2) \label{eq:pd7}\\
&\;\;\; + \mathrm{Pr}(\hat{s}_1= s_1,\hat{s}'_1\neq s'_1,\hat{s}_2 = s_2,\hat{s}'_2= s'_2). \label{eq:pd8}
\end{align}
Now, we consider each of the terms in \eqref{eq:pd1}--\eqref{eq:pd8}. 
\begin{align}
&\mathrm{Pr}(\hat{s}_1\neq s_1) \notag\\
&= \mathrm{Pr}\Big(\exists\hat{s}_1\neq s_1,\tilde{s}'_1,\tilde{s}_2,\tilde{s}'_2, \text{ s.t. } \big(v_1^n(\hat{s}_1),u_1^n(\hat{s}_1,\tilde{s}'_1), \notag\\
&\quad\quad v_2^n(\tilde{s}_2),u_2^n(\tilde{s}_2,\tilde{s}'_2)\big)  \in \calT_\delta^{(n)}(V_1,U_1,V_2,U_2)\notag\\ 
&\quad\quad  \cap \big\{\calB_1(b_1)\times\calB'_1(\hat{s}_1,w_1) \times\calB_2(b_2)\times\calB'_2(\tilde{s}_2,w_2) \big\} \Big)\notag\\
&\leq \mathrm{Pr}\Big( (V_1^n,U_1^n,V_2^n,U_2^n) \in\calT_\delta^{(n)}(V_1,U_1,V_2,U_2)\big| \notag\\
&\quad\quad  (V_1^n,U_1^n)\in\calT_\delta^{(n)}(V_1,U_1),(V_2^n,U_2^n)\in\calT_\delta^{(n)}(V_2,U_2)\Big)\notag\\
&\quad\quad  \cdot 2^{n\sum_{j=1}^2 [I(V_j;Y_j)+\epsilon_1-R_{V_j}+I(U_j;Y_j|V_j)+\epsilon_3-R_{U_j}]}\notag\\
&\leqtop{a} 2^{-n(I(V_1,U_1;V_2,U_2)-\eta)}\notag\\
&\quad\quad  \cdot 2^{n\sum_{j=1}^2 [I(V_j;Y_j)+\epsilon_1-R_{V_j}+I(U_j;Y_j|V_j)+\epsilon_3-R_{U_j}]}\notag\\
&= 2^{n(\sum_{j=1}^2 [I(V_j;Y_j)+\epsilon_1-R_{V_j}+I(U_j;Y_j|V_j)+\epsilon_3-R_{U_j}]-I(U_1;U_2)+\eta)},
\end{align}
where $(a)$ is due to the mutual packing lemma \cite[Lemma~12.2]{ElGamal_NetInformationTheory}. Therefore, $\mathrm{Pr}(\hat{s}_1\neq s_1)$ vanishes as $n\to\infty$ if 
\begin{align}
\sum_{j=1}^2 R_{V_j}+R_{U_j} &\geq \sum_{j=1}^2 I(V_j;Y_j)+I(U_j;Y_j|V_j) - I(U_1;U_2) \notag\\
& \eqtop{a} I(U_1;Y_1)+I(U_2;Y_2)-I(U_1;U_2) \notag\\
& \eqtop{b} I(U_1,U_2;Y_1,Y_2), \label{eq:rv1u1rv2u2}
\end{align}
where $(a)$ and $(b)$ follow from the long Markov chain $V_1-U_1-Y_1-Y_2-U_2-V_2$.

Next, we bound the probability of the event in which $s_1$ is correctly decoded but not $s_2$, i.e., \eqref{eq:pd2} as
\begin{align}
&\mathrm{Pr}(\hat{s}_1=s_1,\hat{s}_2 \neq s_2) \notag\\
&= \mathrm{Pr}\Big(\exists\hat{s}_2\neq s_2,\tilde{s}'_1,\tilde{s}'_2, \text{ s.t. } \big(v_1^n(s_1),u_1^n(s_1,\tilde{s}'_1), \notag\\
&\quad\quad v_2^n(\hat{s}_2),u_2^n(\hat{s}_2,\tilde{s}'_2)\big)  \in \calT_\delta^{(n)}(V_1,U_1,V_2,U_2)\notag\\ 
&\quad\quad  \cap \big\{\calB_1(b_1)\times\calB'_1(s_1,w_1) \times\calB_2(b_2)\times\calB'_2(\hat{s}_2,w_2) \big\} \Big)\notag\\
&\leq \mathrm{Pr}\Big( (V_1^n,U_1^n,V_2^n,U_2^n) \in\calT_\delta^{(n)}(V_1,U_1,V_2,U_2)\big| \notag\\
&\quad\quad  (V_1^n,U_1^n)\in\calT_\delta^{(n)}(V_1,U_1),(V_2^n,U_2^n)\in\calT_\delta^{(n)}(V_2,U_2)\Big)\notag\\
&\quad\quad  \cdot 2^{n(I(V_2;Y_2)+\epsilon_1-R_{V_2}+\sum_{j=1}^2 [I(U_j;Y_j|V_j)+\epsilon_3-R_{U_j}])}\notag\\
&\leq  2^{n(I(V_2;Y_2)+\epsilon_1-R_{V_2}+\sum_{j=1}^2 [I(U_j;Y_j|V_j)+\epsilon_3-R_{U_j}]-I(U_1;U_2)+\eta)},
\end{align}
Therefore, $\mathrm{Pr}(\hat{s}_1=s_1,\hat{s}_2 \neq s_2)$ vanishes as $n\to\infty$ if 
\begin{align}
R_{V_2} + \sum_{j=1}^2 R_{U_j} &\geq I(V_2;Y_2)+ \sum_{j=1}^2 I(U_j;Y_j|V_j) - I(U_1;U_2) \notag\\
& \eqtop{a} I(U_1;Y_1|V_1)+I(U_2;Y_2|U_1) \notag\\
& \eqtop{b} I(U_1;Y_1,Y_2|V_1)+I(U_2;Y_1,Y_2|V_1,U_1) \notag\\
& \eqtop{c} I(U_1,U_2;Y_1,Y_2|V_1), \label{eq:rv2u1u2}
\end{align}
where $(a)$, $(b)$, and $(c)$ follow from the Markov chain $V_1-U_1-Y_1-Y_2-U_2-V_2$. %
With similar steps, $\mathrm{Pr}(\hat{s}_1 \neq s_1,\hat{s}_2=s_2)$ in \eqref{eq:pd3} tends to zero as $n\to\infty$ if 
\begin{align}
R_{V_1} + \sum_{j=1}^2 R_{U_j} \geq  I(U_1,U_2;Y_1,Y_2|V_2). \label{eq:rv1u1u2}
\end{align}
We now bound \eqref{eq:pd4} where $s_1$ and $s_2$ are decoded correctly:
\begin{align}
&\mathrm{Pr}(\hat{s}_1=s_1,\hat{s}'_1\neq s'_1,\hat{s}_2 = s_2) \notag\\
&= \mathrm{Pr}\Big(\exists\hat{s}'_1\neq s'_1,\tilde{s}'_2, \text{ s.t. } \big(v_1^n(s_1),u_1^n(s_1,\hat{s}'_1), \notag\\
&\quad\quad v_2^n(s_2),u_2^n(s_2,\tilde{s}'_2)\big)  \in \calT_\delta^{(n)}(V_1,U_1,V_2,U_2)\notag\\ 
&\quad\quad  \cap \big\{\calB_1(b_1)\times\calB'_1(s_1,w_1) \times\calB_2(b_2)\times\calB'_2(s_2,w_2) \big\} \Big)\notag\\
&= \mathrm{Pr}\Big(\big((V_1^n,U_1^n),V_2^n,U_2^n\big) \in\calT_\delta^{(n)}(V_1,U_1,V_2,U_2) \notag\\
&\quad\quad  \cap (V_1^n,U_1^n,V_2^n) \in\calT_\delta^{(n)}(V_1,U_1,V_2)\notag\\
&\quad\quad  \cap \big\{\calB_1(b_1)\times\calB'_1(s_1,w_1) \times\calB_2(b_2)\times\calB'_2(s_2,w_2) \big\}\Big)\notag\\
&= \mathrm{Pr}\Big(\big((V_1^n,U_1^n),V_2^n,U_2^n\big) \in\calT_\delta^{(n)}(V_1,U_1,V_2,U_2)\big| \notag\\
&\quad\quad  (V_1^n,U_1^n)\in\calT_\delta^{(n)}(V_1,U_1),(V_2^n,U_2^n)\in\calT_\delta^{(n)}(V_2,U_2)\Big) \notag\\
&\quad\quad  \cdot \mathrm{Pr}\Big((V_1^n,U_1^n,V_2^n) \in\calT_\delta^{(n)}(V_1,U_1,V_2)\big| \notag\\
&\quad\quad  (V_1^n,U_1^n)\in\calT_\delta^{(n)}(V_1,U_1),(V_1^n,V_2^n)\in\calT_\delta^{(n)}(V_1,V_2)\Big)\notag\\
&\quad\quad  \cap \big\{\calB_1(b_1)\times\calB'_1(s_1,w_1) \times\calB_2(b_2)\times\calB'_2(s_2,w_2) \big\}\Big)\notag\\
&\leqtop{a} 2^{-n(I(V_1,U_1;U_2|V_2)-\eta)}\cdot 2^{-n(I(U_1;V_2|V_1)-\eta)}\notag\\
&\quad\quad  \cdot 2^{n\sum_{j=1}^2 [I(U_j;Y_j|V_j)+\epsilon_3-R_{U_j}]}\notag\\
&=2^{n(\sum_{j=1}^2 [I(U_j;Y_j|V_j)+\epsilon_3-R_{U_j}]-I(V_1,U_1;U_2|V_2)-I(U_1;V_2|V_1)+2\eta)},
\end{align}
where $(a)$ follows from the joint typicality lemma and the fact that $U_2^n\sim\prod_{i=1}^n p(u_{2,i}|v_{2,i})$, and $U_1^n\sim\prod_{i=1}^n p(u_{1,i}|v_{1,i})$. Therefore, $\mathrm{Pr}(\hat{s}_1=s_1,\hat{s}'_1\neq s'_1,\hat{s}_2 = s_2)$ vanishes as $n\to\infty$ if 
\begin{align}
\sum_{j=1}^2 R_{U_j} &\geq  I(U_1;Y_1|V_1)-I(U_1;V_2|V_1)+I(U_2;Y_2|V_2)\notag\\
&\quad\quad -I(V_1,U_1;U_2|V_2)\notag\\
&\eqtop{a}  I(U_1;Y_1|V_1,V_2)+I(U_2;Y_2|V_1,U_1,V_2)\notag\\
&\eqtop{b}  I(U_1;Y_1,Y_2|V_1,V_2)+I(U_2;Y_1,Y_2|V_1,U_1,V_2)\notag\\
&=  I(U_1,U_2;Y_1,Y_2|V_1,V_2), \label{eq:ru1u2}
\end{align}
where $(a)$ is due the Markov chains $U_1-Y_1-V_2$ and $U_2-Y_1-(U_1,V_1)$, $(b)$ is due to $U_j-Y_j-Y_{j'}$.

Next, for \eqref{eq:pd5}, we have
\begin{align}
&\mathrm{Pr}(\hat{s}_1=s_1,\hat{s}'_1= s'_1,\hat{s}_2 \neq s_2) \notag\\
&= \mathrm{Pr}\Big(\exists\hat{s}_2\neq s_2,\tilde{s}'_2, \text{ s.t. } \big(v_1^n(s_1),u_1^n(s_1,s'_1), \notag\\
&\quad\quad v_2^n(\hat{s}_2),u_2^n(\hat{s}_2,\tilde{s}'_2)\big)  \in \calT_\delta^{(n)}(V_1,U_1,V_2,U_2)\notag\\ 
&\quad\quad  \cap \big\{\calB_1(b_1)\times\calB'_1(s_1,w_1) \times\calB_2(b_2)\times\calB'_2(\hat{s}_2,w_2) \big\} \Big)\notag\\
&\leq \mathrm{Pr}\Big( (V_1^n,U_1^n,V_2^n,U_2^n) \in\calT_\delta^{(n)}(V_1,U_1,V_2,U_2)\big| \notag\\
&\quad\quad  (V_1^n,U_1^n)\in\calT_\delta^{(n)}(V_1,U_1),(V_2^n,U_2^n)\in\calT_\delta^{(n)}(V_2,U_2)\Big)\notag\\
&\quad\quad  \cdot 2^{n(I(V_2;Y_2)+\epsilon_1-R_{V_2}+I(U_2;Y_2|V_2)+\epsilon_3-R_{U_2})}\notag\\
&\leq  2^{n(I(V_2;Y_2)+\epsilon_1-R_{V_2}+I(U_2;Y_2|V_2)+\epsilon_3-R_{U_2}-I(U_1;U_2)+\eta)}.
\end{align}
Therefore, $\mathrm{Pr}(\hat{s}_1=s_1,\hat{s}'_1= s'_1,\hat{s}_2 \neq s_2)$ vanishes as $n\to\infty$ if 
\begin{align}
R_{V_2} + R_{U_2} &\geq I(V_2;Y_2)+ I(U_2;Y_2|V_2) - I(U_1;U_2) \notag\\
& = I(U_2;Y_2|U_1), \label{eq:rv2u2}
\end{align}
which follows from the Markov chain $V_2-U_2-Y_2-U_1$. Similarly,  $\mathrm{Pr}(\hat{s}_1\neq s_1,\hat{s}_2 = s_2,\hat{s}'_2= s'_2)$ in \eqref{eq:pd6} tends to zero as $n\to\infty$ if 
\begin{align}
R_{V_1} + R_{U_1} &\geq I(U_1;Y_1|U_2). \label{eq:rv1u1}
\end{align}
We proceed to bound \eqref{eq:pd7} where $(s_1,s'_1,s_2)$ are decoded correctly:
\begin{align}
&\mathrm{Pr}(\hat{s}_1= s_1,\hat{s}'_1= s'_1,\hat{s}_2 = s_2,\hat{s}'_2\neq s'_2) \notag\\
&= \mathrm{Pr}\Big(\exists\hat{s}'_2\neq s'_2, \text{ s.t. } \big(v_1^n(s_1),u_1^n(s_1,s'_1), \notag\\
&\quad\quad v_2^n(s_2),u_2^n(s_2,hat{s}'_2)\big)  \in \calT_\delta^{(n)}(V_1,U_1,V_2,U_2)\notag\\ 
&\quad\quad  \cap \big\{\calB_1(b_1)\times\calB'_1(s_1,w_1) \times\calB_2(b_2)\times\calB'_2(s_2,w_2) \big\} \Big)\notag\\
&\leq \mathrm{Pr}\Big( (V_1^n,U_1^n,V_2^n,U_2^n) \in\calT_\delta^{(n)}(V_1,U_1,V_2,U_2)\big| \notag\\
&\quad\quad  \big((V_1^n,U_1^n),V_2^n\big)\in\calT_\delta^{(n)}(V_1,U_1,V_2),\notag\\
&\quad\quad  (V_2^n,U_2^n)\in\calT_\delta^{(n)}(V_2,U_2)\Big)\notag\\
&\quad\quad  \cdot 2^{n(I(U_2;Y_2|V_2)+\epsilon_3-R_{U_2})}\notag\\
&\leq  2^{n(I(U_2;Y_2|V_2)+\epsilon_3-R_{U_2}-I(V_1,U_1;U_2|V_2)+\eta)},
\end{align}
from the joint typicality lemma by considering $(V_1,U_1)$ as one variable and the fact that $U_2^n\sim\prod_{i=1}^n p(u_{2,i}|v_{2,i})$. Therefore, $\mathrm{Pr}(\hat{s}_1= s_1,\hat{s}'_1\neq s'_1,\hat{s}_2 = s_2,\hat{s}'_2= s'_2)$ tends to zero as $n\to\infty$ if
\begin{align}
R_{U_2} &\geq I(U_2;Y_2|V_2) - I(V_1,U_1;U_2|V_2) \notag\\
&= I(U_2;Y_2|V_2,U_1), \label{eq:ru2}
\end{align}
where the last equality is due to $V_1-U_1-Y_2-U_2$. Similarly,  $\mathrm{Pr}(\hat{s}_1= s_1,\hat{s}'_1\neq s'_1,\hat{s}_2 = s_2,\hat{s}'_2= s'_2)$ in \eqref{eq:pd8} vanishes as $n\to\infty$ if 
\begin{align}
R_{U_1} &\geq I(U_1;Y_1|V_1,U_2). \label{eq:ru1}
\end{align}
Gathering \eqref{eq:rv1u1rv2u2}--\eqref{eq:ru1}, probability of error event $\mathsf{E}_8$ in \eqref{eq:pd} vanishes as $n\to\infty$ if, for $j,j'=1,2$ and $j\neq j'$, the rates $R_j=R_{V_j}+R_{U_j}$ satisfy 
\begin{align}
R_1  &\geq  I(U_1;Y_1|U_2), \label{eq:conR1}\\
R_2  &\geq  I(U_2;Y_2|U_1),\\
R_1 + R_2  &\geq  I(U_1,U_2;Y_1,Y_2),\\
R_{U_j}&\geq I(U_j;Y_j|V_j,U_{j'}), \label{eq:R_U} \\
R_{U_1} + R_{U_2} &\geq I(U_1,U_2;Y_1,Y_2|V_1,V_2), \label{eq:R_U1U2}\\
R_j + R_{U_{j'}} &\geq I(U_1,U_2;Y_1,Y_2|V_{j'}). \label{eq:R_R1U2}
\end{align}

Finally, $\mathrm{Pr}(\mathsf{E})$ in \eqref{eq:union_bound} tends to zero as $n\to\infty$ provided that \eqref{eq:conR1}--\eqref{eq:R_R1U2} are satisfied.

\emph{\textbf{Equivocation rates}}: %
The equivocation rate with respect to Agent~1, averaged over all codebooks $\calC_n$, is written as
\begin{align}
&H(X^n|f_1(Y_1^n),E^n,\calC_n) = H(X^n|b_1,w_1,E^n,\calC_n) \notag\\
&= H(X^n,Y_1^n|b_1,w_1,E^n,\calC_n)-H(Y_1^n|b_1,w_1,E^n,X^n,\calC_n)\notag\\
&= H(Y_1^n|b_1,w_1,E^n,\calC_n)+H(X^n|b_1,w_1,E^n,Y_1^n,\calC_n) \notag\\
&\qquad -H(Y_1^n|b_1,w_1,E^n,X^n,\calC_n)\notag\\
&= H(Y_1^n|b_1,w_1,E^n,\calC_n)+H(X^n|E^n,Y_1^n)\notag\\
&\qquad  -H(Y_1^n|b_1,w_1,X^n,\calC_n) \label{eq:equiv_proof}
\end{align}
where the last equality holds since $f_1(Y_1^n)=(b_1,w_1)$ is a deterministic function, $Y_1^n-(b_1,w_1,X^n,\calC_n)-E^n$ is a Markov chain, and $(X^n,Y_1^n,E^n)$ are independent of the codebook. The first term in \eqref{eq:equiv_proof} is bounded as
\begin{align}
H(&Y_1^n|b_1,w_1,E^n,\calC_n) \notag\\
&= H(Y_1^n|b_1,E^n,\calC_n)-I(Y_1^n;w_1|b_1,E^n,\calC_n)\notag\\
&\geqtop{a} H(Y_1^n|b_1,E^n,\calC_n)-H(w_1|b_1,E^n,\calC_n)\notag\\
&\geqtop{b} H(Y_1^n|s_1,E^n,\calC_n)-H(w_1|\calC_n)\notag\\
&= H(Y_1^n,E^n|s_1,\calC_n)-H(E^n|s_1,\calC_n)-H(w_1|\calC_n)\notag\\
&\eqtop{c} H(Y_1^n,E^n)-H(s_1|\calC_n)+H(s_1|Y_1^n,E^n,\calC_n)\notag\\
&\qquad -H(E^n|s_1,\calC_n)-H(w_1|\calC_n)\notag\\
&\geqtop{d} H(Y_1^n,E^n)-H(s_1|\calC_n)-H(E^n|s_1,\calC_n)-H(w_1|\calC_n)\notag\\
&\geqtop{e} nH(Y_1,E)-nI(V_1;Y_1)-n\epsilon_1-H(E^n|s_1,\calC_n) \notag\\
&\qquad -nR_{U_1}-n\epsilon_4\notag\\
&\geqtop{f} n\big[H(Y_1,E)-I(V_1;Y_1)-H(E|V_1)-R_{U_1}-\epsilon'\notag\\
&\qquad -\epsilon_1-\epsilon_4\big]\notag\\
&\geqtop{g} n\big[H(Y_1|E)-I(V_1;Y_1|E)-R_{U_1}-\epsilon'-\epsilon_1-\epsilon_4\big]\notag\\
&= n\big[H(Y_1|V_1E)-R_{U_1}-\epsilon'-\epsilon_1-\epsilon_4\big], \label{eq:equiv1}
\end{align}
where
\begin{enumerate}[(a)]
\item follows as the bin index $w_1$ is a deterministic function of $Y_1^n$;
\item follows since the bin index $b_1$ is a deterministic function of the codeword index $s_1$, and conditioning reduces entropy;
\item follows since $(Y_1^n,E^n)$ are independent of the codebook;
\item follows since the codeword index $s_1$ is a deterministic function of $Y_1^n$;
\item follows since random variables $Y_{1,i}$ and $E_i$ are i.i.d., $s_1$ and $w_1$ are random variables over sets of size $2^{n(I(V_1;Y_1)+\epsilon_1)}$ and $2^{n(R_{U_1}+\epsilon_4)}$, respectively;
\item follows from \eqref{eq:lem1} in Lemma~\ref{lem:inner};
\item follows from the Markov chain $V_1-Y_1-E$.
\end{enumerate}
The last term in \eqref{eq:equiv_proof} can be bounded as
\begin{align}
H(&Y_1^n|b_1,w_1,X^n,\calC_n) \eqtop{a} H(Y_1^n|b_1,w_1,U_2^n,X^n,\calC_n) \notag\\
&= H(Y_1^n|b_1,w_1,U_1^n,U_2^n,X^n,\calC_n)\notag\\
&\qquad +I(U_1^n;Y_1^n|b_1,w_1,U_2^n,X^n,\calC_n)\notag\\
&\leqtop{b}  H(Y_1^n|U_1^n,X^n,\calC_n)+H(U_1^n|b_1,w_1,U_2^n,X^n,\calC_n)\notag\\
&\leqtop{c}  n\big[H(Y_1|U_1,X) + \epsilon\big], \label{eq:equiv2}
\end{align}
where
\begin{enumerate}[(a)]
\item follows from the Markov chain $Y_1^n\!-\!(b_1,w_1,X^n,\calC_n)\!-\!U_2^n$;
\item follows since conditioning reduces entropy, the bin index $w_1$ is a deterministic function of $U_1^n$, and the Markov chain $U_1^n-(X^n,\calC_n)-U_2^n$;
\item follows from Fano's inequality (i.e., in the decoding scheme having $U_2^n$, $b_1$, and $w_1$, the CEO decodes $U_1^n$ with high probability). 
\end{enumerate}
Substituting \eqref{eq:equiv1} and \eqref{eq:equiv2} in \eqref{eq:equiv_proof}, we obtain
\begin{align}
H\big(&X^n|f_1(Y_1^n),E^n,\calC_n\big) \notag\\
&\geq n\big[H(Y_1|V_1,E)-R_{U_1}+H(X|Y_1,E)-H(Y_1|U_1,X)\notag\\
&\qquad -\epsilon-\epsilon'-\epsilon_1-\epsilon_4\big].
\end{align}
Therefore,
\begin{equation}
\Delta_1  \leq  H(Y_1|V_1,E)-R_{U_1}+H(X|Y_1,E)-H(Y_1|U_1,X)
\end{equation}
is achievable. Based on \eqref{eq:R_U}, using Fourier-Motzkin elimination yields
\begin{align}
\Delta_1  &\leq  H(Y_1|V_1,E)-I(U_1;Y_1|V_1,U_2)+H(X|Y_1,E)\notag\\
&\qquad -H(Y_1|U_1,X)\notag\\
&\eqtop{a} H(Y_1|V_1,E)-I(U_1;Y_1|V_1,U_2)+H(X|Y_1,E,V_1)\notag\\
&\qquad -H(Y_1|U_1,V_1,X)\notag\\
&= H(Y_1,X|V_1,E)-I(U_1;Y_1|V_1,U_2)\notag\\
&\qquad -H(Y_1|U_1,V_1,X)\notag\\
&= H(X|V_1,E)+H(Y_1|V_1,X,E)-I(U_1;Y_1|V_1,U_2)\notag\\
&\qquad -H(Y_1|U_1,V_1,X)\notag\\
&\eqtop{b} H(X|V_1,E)-I(U_1;Y_1|V_1,U_2)+I(U_1;Y_1|V_1,X),
\end{align}
where
\begin{enumerate}[(a)]
\item follows from the Markov chains $V_1-(U_1,X)-Y_1$ and $V_1-(Y_1,E)-X$;
\item follows from the Markov chain $Y_1-(V_1,X)-E$.
\end{enumerate}

Similarly for the second link, we obtain
\begin{equation}
\Delta_2  \leq  H(X|V_2,E)- I(U_2;Y_2|V_2,U_1) + I(U_2;Y_2|V_2,X).
\end{equation}

For the sum of the equivocation rates, following the proof of the equivocation rate, we have
\begin{align}
\Delta_1  +  \Delta_2 &\leq H(X|V_1,E)+H(X|V_2,E)-R_{U_1}-R_{U_2}\notag\\
&\qquad +I(U_1;Y_1|V_1,X) +I(U_2;Y_2|V_2,X)\notag\\
&\leq H(X|V_1,E)+H(X|V_2,E) \notag\\
&\qquad - I(U_1,U_2;Y_1,Y_2|V_1,V_2) \notag\\
&\qquad + I(U_1;Y_1|V_1,X) + I(U_2;Y_2|V_2,X),
\end{align}
where we substitute \eqref{eq:R_U1U2} in the second inequality. In order to prove \eqref{eq:delta1R2_inner}, we have
\begin{align}
\Delta_1  &\leq  H(X|V_1,E) - R_{U_1} + I(U_1;Y_1|V_1,X)\notag\\
&\leqtop{a}  H(X|V_1,E) - I(U_1,U_2;Y_1,Y_2|V_1) + R_2 \notag\\
&\qquad + I(U_1;Y_1|V_1,X)\notag\\
&\eqtop{b} H(X|V_1,E) - I(U_1;Y_1|V_1) - I(U_2;Y_2|V_1,U_1) \notag\\
&\qquad + R_2  + I(U_1;Y_1|V_1,X)\notag\\
&\eqtop{c} H(X|V_1,E) - I(U_1;Y_1|V_1) - I(U_2;Y_2|U_1) + R_2 \notag\\
&\qquad  + I(U_1;Y_1|V_1,X),
\end{align}
where
\begin{enumerate}[(a)]
\item follows from \eqref{eq:R_R1U2};
\item follows since $U_1-Y_1-Y_2-U_2$ is a Markov chain;
\item follows since $V_1-U_1-U_2$ forms a Markov chain.
\end{enumerate}
Inequality \eqref{eq:delta2R1_inner} can be proved similarly.

\emph{\textbf{Distortion at the CEO}}: The distortion at the CEO is written as
\begin{align}
\E\Big[&d\Big(X^n,g(f_1(Y_1^n),f_2(Y_2^n))\Big)\Big] \leq \mathrm{Pr}(\mathsf{E})d_{\max} \notag\\
&\qquad + \mathrm{Pr}(\mathsf{E}^c) \E\Big[d\Big(X^n,\hat{X}\big(u_{1}^n(s_1,s'_1),u_{2}^n(s_2,s'_2)\big)\Big)\Big|\mathsf{E}^c \Big] \notag\\
&\leqtop{a} \epsilon_n \!+\! \frac{1}{n}\sum_{i=1}^n \E\Big[d\Big(X_i,\hat{X}\big(u_{1,i}(s_1,s'_1),u_{2,i}(s_2,s'_2)\big)\Big)\Big|\mathsf{E}^c \Big] \notag\\
&\leqtop{b} \epsilon_n + \E\Big[d\big(X,\hat{X}(U_1,U_2)\big)\Big],
\end{align}
where
\begin{enumerate}[(a)]
\item follows since the probability of an error event becomes small as $n\to\infty$ if the rate constraints are satisfied.
\item follows since for every jointly typical sequences $(x^n,u_1^n,u_2^n)$, we have 
\begin{align}
d&\big(x^n,\hat{X}(u_1^n,u_2^n)\big)=\frac{1}{n}\sum_{i=1}^n d\big(x_i,\hat{X}(u_{1,i},u_{2,i})\big)\notag\\
&= \frac{1}{n}\sum_{(x,u_1,u_2)\in\calX \times \calU_1 \times \calU_2} \Big[d\big(x,\hat{X}(u_{1},u_{2})\big)\notag\\
&\qquad\qquad\qquad\qquad\qquad\qquad N\big(x,u_1,u_2|x^n,u_1^n,u_2^n\big)\Big]\notag\\
&= \E\Big[d\big(X,\hat{X}(U_1,U_2)\big)\Big] \notag\\
& \quad \quad + \frac{1}{n}\sum_{(x,u_1,u_2)\in\calX \times \calU_1 \times \calU_2} \Big[d\big(x,\hat{X}(u_{1},u_{2})\big) \notag\\
&\qquad\qquad \Big(\frac{1}{n} N\big(x,u_1,u_2|x^n,u_1^n,u_2^n\big)-p(x,u_1,u_2)\Big)\Big]\notag\\
&\leq \E\Big[d\big(X,\hat{X}(U_1,U_2)\big)\Big] + d_{\max}|\calX||\calU_1||\calU_2|\epsilon'_n,
\end{align}
\end{enumerate}
with $\epsilon'_n\to 0$ as $n\to\infty$ from properties of jointly (strong) typical sequences. Therefore, to satisfy the distortion constraint at the CEO, it is sufficient to have
\begin{equation}
\E\Big[d\big(X,\hat{X}(U_1,U_2)\big)\Big] \leq D.
\end{equation}
This concludes the proof of Theorem~\ref{th:inner}.\QEDA


\section{Proof of Proposition~\ref{prop:card_inner}: Cardinalities in the Inner Bound}
\label{sec:proof_card_inner}
To bound the cardinality of alphabets of auxiliary random variables $V_1$ and $U_1$ in Theorem~\ref{th:inner}, we rewrite the equations of the inner bound using  straightforward derivations as
\begin{align}
R_1  &\geq  H(Y_1|U_2) - H(Y_1|U_1,U_2), \\
R_2  &\geq  I(U_2;Y_2|U_1),\\
R_1 + R_2  &\geq  H(Y_1) - H(Y_1|U_1) + I(U_2;Y_2|U_1), \\
\Delta_1  &\leq  \big[I(X;U_2|V_1) - I(X;E|V_1) + H(X|U_1,U_2)\big]^+, \\
\Delta_2  &\leq  \big[H(X|V_2,E)\! -\! H(X|V_2,U_1)\! +\! H(X|U_1,U_2)\big]^+, \\
\Delta_1  +  \Delta_2  &\leq  \big[I(X;V_2|V_1) - I(X;E|V_1)  \notag\\
&\qquad +  H(X|V_2,E) + H(X|U_1,U_2)\big]^+, \\
\Delta_1 - R_2  &\leq  \big[I(X;U_2|V_1) - I(X;E|V_1) \notag\\
&\qquad - I(U_2;Y_2|V_1) + H(X|U_1,U_2)\big]^+,  \\
\Delta_2 - R_1  &\leq  \big[H(X|V_2,E) - H(Y_1|U_2)  \notag\\
&\qquad + H(Y_1|U_1,U_2) -  I(U_2;X|V_2) \big]^+,  \\
D&\geq\E\Big[d\big(X,\hat{X}(U_1,U_2)\big)\Big].
\end{align}
Using standard arguments based on the Fenchel-Eggleston-Carath\'{e}odory theorem and the support lemma \cite[Appendix~C]{ElGamal_NetInformationTheory}, it can be shown that $\calV_1$ should have $|\calY_1|-1$ letters to preserve the probability distribution $p(y_1)$, and ten more to preserve $H(Y_1|U_1,U_2)$, $I(U_2;Y_2|U_1)$, $H(Y_1|U_1)$, $I(X;U_2|V_1)$, $I(X;E|V_1)$, $H(X|U_1,U_2)$, $H(X|V_2,U_1)$, $I(X;V_2|V_1)$, $I(U_2;Y_2|V_1)$, and the distortion constraint. Furthermore, for each $v_1\in\calV_1$, there exists a set $\calU_1'$ with $|\calY_1|-1$ letters to preserve  the distribution $p(y_1)$, plus six more to preserve $H(Y_1|U_1,U_2)$, $I(U_2;Y_2|U_1)$, $H(Y_1|U_1)$, $H(X|U_1,U_2)$, $H(X|V_2,U_1)$, and the distortion constraint.

Therefore, it suffices to have $|\calV_1|\leq |\calY_1|+9$ and $|\calU_1|\leq(|\calY_1|+9)(|\calY_1|+5)$. The same holds for the cardinalities of the sets $\calV_2$ and $\calU_2$.\QEDA


\section{Proof of Theorem~\ref{th:outer}: The Outer Bound}
\label{sec:proof_outer}
We denote the message transmitted by the $j$th agent as $J_j=f_j(Y_j^n)$, and also define auxiliary random variables
\begin{align}
V_{j,i} \coloneqq\, &(J_j,X^{i-1}), \label{eq:auxV}\\
U_{j,i} \coloneqq\, &(V_{j,i},Y_1^{i-1},Y_2^{i-1}) \nonumber\\
=\, &(J_j,X^{i-1},Y_1^{i-1},Y_2^{i-1}), \label{eq:auxU}
\end{align}
for $j\in\{1,2\}$ and $i\in\{1,\dots,n\}$. %
Then, we have the following chain of inequalities for Agent~1's rate:
\begin{align}
n&(R_1+\epsilon) \geq H(J_1)\notag \\
&\eqtop{a} I(J_1;J_2,X^n,Y_1^n,Y_2^n) \notag\\
&\geqtop{b} I(J_1;X^n,Y_1^n,Y_2^n|J_2) \notag\\
&\eqtop{c} \sum_{i=1}^n I(J_1;X_i,Y_{1,i},Y_{2,i}|J_2,X^{i-1},Y_1^{i-1},Y_2^{i-1}) \notag\\
&\geqtop{d} \sum_{i=1}^n I(J_1;Y_{1,i}|J_2,X^{i-1},Y_1^{i-1},Y_2^{i-1}) \notag\\
&= \sum_{i=1}^n I(J_1,X^{i-1},Y_1^{i-1},Y_2^{i-1};Y_{1,i}|J_2,X^{i-1},Y_1^{i-1},Y_2^{i-1}) \notag\\
&\eqtop{e} \sum_{i=1}^n I(U_{1,i};Y_{1,i}|U_{2,i}),
\end{align}
where
\begin{enumerate}[(a)]
\item follows as $J_1=f_1(Y_1^n)$;
\item follows from the non-negativity of mutual information;
\item follows from the chain rule of conditional mutual information;
\item follows from the non-negativity of mutual information;
\item follows from the definition of auxiliary random variables in \eqref{eq:auxV}--\eqref{eq:auxU}.
\end{enumerate}
The equivocation with respect to Agent~1 is bounded as
\begin{align}
n(&\Delta_1-\epsilon) \leq H(X^n|J_1,E^n) \notag\\
&= H(X^n|J_1)- I(X^n;E^n|J_1) \notag\\
&\eqtop{a} H(X^n|J_1) - I(X^n;E^n) + I(J_1;E^n) \notag\\
&\eqtop{b} \sum_{i=1}^n \Big[H(X_i|J_1,X^{i-1}) - I(X_i;E_i) + I(J_1,E^{i-1};E_i)\Big]\notag\\
&\leqtop{c} \sum_{i=1}^n \Big[H(X_i|J_1,X^{i-1}) + I(J_1,X^{i-1},E^{i-1};E_i) \notag\\
&\qquad\quad  - I(X_i;E_i)\Big]\notag\\
&\eqtop{d} \sum_{i=1}^n \Big[H(X_i|J_1,X^{i-1}) - I(X_i;E_i|J_1,X^{i-1}) \Big]\notag\\
&\eqtop{e} \sum_{i=1}^n \Big[ H(X_i|V_{1,i}) - I(X_i;E_i|V_{1,i}) \Big]\notag\\
&= \sum_{i=1}^n \Big[H(X_i|E_i) - I(X_i;V_{1,i}|E_i) \Big]
\end{align}
where
\begin{enumerate}[(a)]
\item follows since $J_1-X^n-E^n$ is a Markov chain;
\item follows since $X_i$ and $E_i$ are memoryless, and the chain rule of conditional entropy;
\item follows the non-negativity of mutual information;
\item follows from the Markov chain $E_i\!-\!X_i\!-\!(J_1,X^{i-1})\!-\!E^{i-1}$;
\item follows from the definition of auxiliary random variable in \eqref{eq:auxV}. 
\end{enumerate}

To prove \eqref{eq:delta1R2_outer}, we have
\begin{align}
n&(\Delta_1-\epsilon) \leq H(X^n|J_1,E^n) \notag\\
&= H(X^n|J_1,J_2,E^n) + I(X^n;J_2|J_1,E^n) \notag\\
&\leqtop{a} \sum_{i=1}^n \Big[H(X_i|J_1,J_2,X^{i-1},E^n)\Big] + H(J_2)\notag\\
&\leqtop{b} \sum_{i=1}^n \Big[H(X_i|J_1,J_2,X^{i-1},E_i)\Big] + H(J_2)\notag\\
&\eqtop{c} \sum_{i=1}^n \Big[H(X_i|V_{1,i},V_{2,i},E_i)\Big] + H(J_2)\notag\\
&\leq \sum_{i=1}^n \Big[H(X_i|V_{1,i},E_i)-I(X_i;V_{2,i}|V_{1,i},E_i)\Big] + nR_2 + n\epsilon,
\end{align}
where
\begin{enumerate}[(a)]
\item follows from the non-negativity of conditional mutual information and the fact that conditioning reduces entropy;
\item follows since conditioning reduces entropy:
\item follows from definition of auxiliary random variables \eqref{eq:auxV}.
\end{enumerate}

Since the setup is symmetric, bounds on Agent~2's transmission and equivocation rates in \eqref{eq:R2_outer}, \eqref{eq:delta2_outer}, and \eqref{eq:delta2R1_outer} are derived similar to those of Agent~1. %
The sum rate is bounded as
\begin{align}
n(&R_1+R_2+2\epsilon) \geq H(J_1,J_2) \notag\\
&\eqtop{a} I(J_1,J_2;X^n,Y_1^n,Y_2^n) \notag\\
&\eqtop{b} \sum_{i=1}^n I(J_1,J_2;X_i,Y_{1,i},Y_{2,i}|X^{i-1},Y_1^{i-1},Y_2^{i-1}) \notag\\
&= \sum_{i=1}^n I(J_1,J_2,X^{i-1},Y_1^{i-1},Y_2^{i-1};X_i,Y_{1,i},Y_{2,i}) \notag\\
&\qquad\quad -I(X^{i-1},Y_1^{i-1},Y_2^{i-1};X_i,Y_{1,i},Y_{2,i}) \notag\\
&\eqtop{c} \sum_{i=1}^n I(J_1,J_2,X^{i-1},Y_1^{i-1},Y_2^{i-1};X_i,Y_{1,i},Y_{2,i})\notag\\
&\geqtop{d} \sum_{i=1}^n I(J_1,J_2,X^{i-1},Y_1^{i-1},Y_2^{i-1};Y_{1,i},Y_{2,i})\notag\\
&\eqtop{e} \sum_{i=1}^n I(U_{1,i},U_{2,i};Y_{1,i},Y_{2,i}),
\end{align}
where
\begin{enumerate}[(a)]
\item follows as $J_1=f_1(Y_1^n)$ and $J_2=f_2(Y_2^n)$;
\item follows from the chain rule of mutual information;
\item follows since $X_i,Y_{1,i},Y_{2,i}$ are memoryless;
\item follows from the non-negativity of mutual information;
\item follows from definition of auxiliary random variables \eqref{eq:auxU}.
\end{enumerate}
Next, the distortion bound at the CEO is written as
\begin{align}
D+\epsilon &\geq \E\Big[d\big(X^n,g(J_1,J_2)\big)\Big] \notag\\
&\geq \frac{1}{n}\sum_{i=1}^n \E\Big[d\Big(X_i,\hat{X}_i\big(U_{1,i},U_{2,i}\big)\Big)\Big],
\end{align}
where $\hat{X}_i(U_{1,i},U_{2,i})\coloneqq g_i(J_1,J_2)$ is the $i$th element of the CEO's decoded sequence.

We define a time sharing random variable $Q$ independent from all other random variables and uniformly distributed over the set $\{1,\dots,n\}$. We further define $X=X_Q$, $Y_j=Y_{j,Q}$, $E=E_Q$, $U_j=(Q,U_{j,Q})$, and $V_j=(Q,V_{j,Q})$ for $j\in\{1,2\}$. Also let  $\hat{X}(U_1,U_2)=\hat{X}(Q,U_{1,Q},U_{2,Q})=\hat{X}_Q(U_{1,Q},U_{2,Q})$. We have the Markov chains $V_j-U_j-Y_j-(X,E,Y_{j'})$ for $j,j'\in\{1,2\}$ and $j\neq j'$. Now, we can write the rates and equivocation bounds for Agent~1 (and similarly for Agent~2) as follows:  
\begin{align}
R_1+\epsilon &\geq \frac{1}{n} \sum_{i=1}^n I(U_{1,i};Y_{1,i}|U_{2,i}) \notag\\
&= \frac{1}{n} \sum_{i=1}^n I(U_{1,Q};Y_{1,Q}|U_{2,Q},Q=i)\notag\\
&= I(U_{1};Y_{1}|U_{2}),\\[3pt]
\Delta_1-\epsilon &\leq \frac{1}{n} \sum_{i=1}^n \Big[H(X_i|E_i) - I(X_i;V_{1,i}|E_i) \Big]\notag\\
&= \frac{1}{n} \sum_{i=1}^n \Big[H(X_Q|E_Q,Q=i) \notag\\
&\qquad\qquad - I(X_Q;V_{1,Q}|E_Q,Q=i) \Big]\notag\\
&=  H(X|E) - I(X;V_{1}|E),\\[3pt]
\Delta_1-R_2&-2\epsilon \notag\\
&\leq \frac{1}{n} \sum_{i=1}^n \Big[H(X_i|V_{1,i},E_i) -I(X_i;V_{2,i}|V_{1,i},E_i)\Big]\notag\\
&= \frac{1}{n} \sum_{i=1}^n \Big[H(X_Q|V_{1,Q},E_Q,Q=i) \notag\\
&\qquad\qquad -I(X_Q;V_{2,Q}|V_{1,Q},E_Q,Q=i)\Big]\notag\\
&= H(X|V_{1},E) -I(X;V_{2}|V_{1},E).
\end{align}
The sum rate and the distortion bounds can also be written as
\begin{align}
R_1+R_2+2\epsilon &\geq \frac{1}{n} \sum_{i=1}^n I(U_{1,i},U_{2,i};Y_{1,i},Y_{2,i})\notag\\
&=\frac{1}{n} \sum_{i=1}^n I(U_{1,Q},U_{2,Q};Y_{1,Q},Y_{2,Q}|Q=i)\notag\\
&=I(U_{1},U_{2};Y_{1},Y_{2}),\\[3pt]
D+\epsilon &\geq \frac{1}{n}\sum_{i=1}^n \E\Big[d\Big(X_i,\hat{X}_i\big(U_{1,i},U_{2,i}\big)\Big)\Big]\notag\\
&= \frac{1}{n}\sum_{i=1}^n \E\Big[d\Big(X_Q,\hat{X}_Q\big(U_{1,Q},U_{2,Q}\big)\Big)\Big|Q=i\Big]\notag\\
&= \E\Big[d\Big(X_Q,\hat{X}_Q\big(U_{1,Q},U_{2,Q}\big)\Big)\Big]\notag\\
&= \E\Big[d\Big(X,\hat{X}\big(U_{1},U_{2}\big)\Big)\Big].
\end{align}
Letting $\epsilon$ and $\epsilon_n$ tend to zero, inequalities of the Theorem~\ref{th:outer} follow.\QEDA


\section{Proof of Proposition~\ref{prop:card_outer}: Cardinalities in the Outer Bound}
\label{sec:proof_card_outer}

To bound the cardinality of the sets $\calV_1$ and $\calU_1$ in Theorem~\ref{th:outer}, we rewrite the equations of the outer bound as
\begin{align}
R_1  &\geq  H(Y_1|U_2) - H(Y_1|U_1,U_2), \\
R_2  &\geq  I(U_2;Y_2|U_1),\\
R_1 + R_2  &\geq  H(Y_1) - H(Y_1|U_1) + I(U_2;Y_2|U_1), \\
\Delta_1  &\leq  H(X|V_1) - I(X;E|V_1), \\
\Delta_2  &\leq  H(X|V_2) - I(X;E|V_2),\\
\Delta_1  - R_2  &\leq  H(X|V_1) - I(X;E|V_1) -I(X;V_2|V_1)  \notag\\
&\qquad  + I(E;V_2|V_1),  \\
\Delta_2 - R_1  &\leq  H(X|V_1) -I(X;V_2|V_1) - I(X;E|V_1)\notag\\
&\qquad  + I(E;V_2|V_1),  \\
D&\geq\E\Big[d\big(X,\hat{X}(U_1,U_2)\big)\Big].
\end{align}
Similar to the approach in Appendix~\ref{sec:proof_card_inner}, using standard arguments based on the Fenchel-Eggleston-Carath\'{e}odory theorem and the support lemma \cite[Appendix~C]{ElGamal_NetInformationTheory}, it can be shown that $\calV_1$ should have $|\calY_1|-1$ letters to preserve the probability distribution $p(y_1)$, and eight more to preserve $H(Y_1|U_1,U_2)$, $I(U_2;Y_2|U_1)$, $H(Y_1|U_1)$, $H(X|V_1)$, $I(X;E|V_1)$, $I(X;V_2|V_1)$, $I(E;V_2|V_1)$, and the distortion constraint. Furthermore, for each $v_1\in\calV_1$, there exists a set $\calU_1'$ with $|\calY_1|-1$ letters to preserve the distribution $p(y_1)$, plus four more to preserve $H(Y_1|U_1,U_2)$, $I(U_2;Y_2|U_1)$, $H(Y_1|U_1)$, and the distortion constraint.

Therefore, it suffices to have $|\calV_1|\leq |\calY_1|+7$ and $|\calU_1|\leq(|\calY_1|+7)(|\calY_1|+3)$. The same holds for the cardinalities of the sets $\calV_2$ and $\calU_2$.\QEDA


\section{Converse Proof of Corollary~\ref{cor:no_Eve_SI_lossless}}
\label{sec:proof_no_Eve_SI}
Let $J_1=f_1(X^n)$ and $J_2=f_2(Y_2^n)$ denote the messages transmitted by Agent~1 and Agent~2, respectively, and also define the auxiliary random variable $U_{2,i} = (J_2,X^{i-1})$, for $i\in\{1,\dots,n\}$.

For the equivocation rates, we have
\begin{align}
n(\Delta_1-\epsilon) &\leq H(X^n|J_1) \notag\\
&= H(X^n,J_2|J_1) - H(J_2|J_1X^n)\notag\\
&= H(J_2|J_1) - H(J_2|J_1X^n) + H(X^n|J_1J_2)\notag\\
&\leqtop{a} H(J_2) - H(J_2|X^n) + n\epsilon_n\notag\\
&= I(J_2;X^n) + n\epsilon_n\notag\\
&= \sum_{i=1}^n I(J_2X^{i-1};X_i) + n\epsilon_n\notag\\
&= \sum_{i=1}^n I(U_{2,i};X_i) + n\epsilon_n,
\end{align}
where (a) follows from Fano's inequality, the Markov chain $J_1-X^n-J_2$, and the fact that conditioning reduces entropy. We also have
\begin{align}
n(\Delta_2-\epsilon) &\leq H(X^n|J_2) \notag\\
&= \sum_{i=1}^n H(X_i|J_2X^{i-1})\notag\\
&= \sum_{i=1}^n H(X_i|U_{2,i}).
\end{align}

Similar to the proof of Theorem~\ref{th:outer}, by using an independent random variable $Q$, inequalities of Corollary~\ref{cor:no_Eve_SI_lossless} follow.\QEDA


\section{Proof of Theorem~\ref{th:gaussian}: Quadratic Gaussian Case without Side Information at Eve}
\label{sec:proof_gauss}

\subsection{Achievability}
\label{sec:proof_gauss_ach}
The achievability is proved by applying the results of Theorem~\ref{th:inner} to the Gaussian case. Although Theorem~\ref{th:inner} is proved for finite alphabet sources with bounded distortion measure, its results can be applied to the Gaussian sources with quadratic distortion measure as shown e.g., in  \cite{Oohama-1997-TIT}. In addition similar results are proved for the Gaussian CEO problem (without secrecy constraints) in \cite{Prabhakaran-2004-ISIT} and \cite{Oohama-2005-TIT}. %

It follows from Theorem~\ref{th:inner} that with no side information at Eve the following region is achievable:
\begin{align}
R_1  &\geq  I(U_1;Y_1|U_2),  \label{eq:R1_gaussian}\\
R_2  &\geq  I(U_2;Y_2|U_1), \label{eq:R2_gaussian}\\
R_1 + R_2  &\geq  I(U_1,U_2;Y_1,Y_2), \label{eq:Rsum_gaussian}\\
\Delta_1  &\leq  h(X) - I(U_1;Y_1|U_2) + I(U_1;Y_1|X),  \label{eq:delta1_gaussian}\\
\Delta_2  &\leq  h(X) - I(U_2;Y_2|U_1) + I(U_2;Y_2|X),  \label{eq:delta2_gaussian}\\
\Delta_1  +  \Delta_2  &\leq  2h(X) - I(U_1,U_2;Y_1,Y_2) \notag\\
&\qquad + I(U_1;Y_1|X) + I(U_2;Y_2|X), \label{eq:delta12_gaussian} \\
\Delta_1  - R_2  &\leq  h(X) -I(U_2;Y_2|U_1) -  I(U_1;Y_1) \notag\\
&\qquad+  I(U_1;Y_1|X), \label{eq:delta1R2_gaussian} \\
\Delta_2 - R_1  &\leq  h(X) - I(U_1;Y_1|U_2) -  I(U_2;Y_2) \notag\\
&\qquad + I(U_2;Y_2|X),  \label{eq:delta2R1_gaussian} \\
D&\geq\E\Big[d\big(X,\hat{X}(U_1,U_2)\big)\Big], \label{eq:D_gaussian}
\end{align}
where $h(\cdot)$ is the differential entropy. This region is obtained by setting $V_1$, $V_2$, and $E$ to be constant in \eqref{eq:delta1_inner}--\eqref{eq:delta2R1_inner}.

Let $U_1$ and $U_2$ be auxiliary random variables jointly distributed with the source $X$ and the respective agent's observation $Y_1$ and $Y_2$ such that
\begin{align}
U_j &= Y_j + Z_j, \quad j\in\{1,2\}\label{eq:Uj_gaussian} 
\end{align}
where $Z_j\sim\calN(0,\sigma_{Z_j}^2)$ for $j\in\{1,2\}$ is independent of $X$, $Y_1$, and $Y_2$. Given $U_1$ and $U_2$,  we choose the estimator function $\hat{X}(U_1,U_2)$ to be the minimum mean square error (MMSE) estimator. For a fixed target distortion $D>0$, the auxiliary random variables are adjusted so that the distortion constraint \eqref{eq:D_gaussian} is satisfied. Therefore, we have
\begin{align}
D &= \E\Big[d\big(X,\hat{X}(U_1,U_2)\big)\Big] = \E\Big[\big(X-\hat{X}(U_1,U_2)\big)^2\Big] \notag\\
&= \frac{1}{2\pi e}2^{2h(X|U_1,U_2)},  \label{eq:est_err}
\end{align}
where \eqref{eq:est_err} follows from \cite[Theorem 8.6.6]{Cover_InformationTheory}. We then calculate
\begin{align}
h(X|U_1,U_2) &= h(X,U_1,U_2) - h(U_1,U_2) \notag\\
&= \frac{1}{2}\log\left((2\pi e)\frac{\mathrm{det}\bsK_{XU_1U_2}}{\mathrm{det}\bsK_{U_1U_2}}\right), \label{eq:cond_ent}
\end{align}
where the covariance matrix $\bsK_{XU_1U_2}$ is
\begin{equation*}
\bsK_{XU_1U_2} =
\begin{bmatrix}
\sigma_X^2 & \sigma_X^2 & \sigma_X^2 \\
\sigma_X^2 & \sigma_X^2+\sigma_{N_1}^2+\sigma_{Z_1}^2 & \sigma_X^2 \\
\sigma_X^2 & \sigma_X^2 & \sigma_X^2+\sigma_{N_2}^2+\sigma_{Z_2}^2
\end{bmatrix}.
\end{equation*}
Substituting \eqref{eq:cond_ent} in \eqref{eq:est_err}, we obtain
\begin{equation}
\label{eq:mmse_D}
\frac{1}{D} = \frac{1}{\sigma_X^2}+\frac{1}{\sigma_{N_1}^2+\sigma_{Z_1}^2}+\frac{1}{\sigma_{N_2}^2+\sigma_{Z_2}^2}.
\end{equation}
Next, for $j\in\{1,2\}$, we define
\begin{align}
r_j &\coloneqq I(U_j;Y_j|X) \label{eq:r_j_def} \\
&= I(U_j;Y_j)-I(U_j;X) \\
&= \frac{1}{2}\log\frac{\sigma_{N_j}^2+\sigma_{Z_j}^2}{\sigma_{Z_j}^2}, \label{eq:r_j}
\end{align}
where the second equality is due to the Markov chain $U_j-Y_j-X$. For any $r_j\geq 0$ with $j\in\{1,2\}$, there exists corresponding $\sigma_{Z_j}^2$, and thus, auxiliary random variable $U_j$. Then, using  \eqref{eq:r_j}, we can rewrite \eqref{eq:mmse_D} as
\begin{equation}
\label{eq:mmse_D_rj}
\frac{1}{D} = \frac{1}{\sigma_X^2}+\frac{1-2^{-2r_1}}{\sigma_{N_1}^2}+\frac{1-2^{-2r_2}}{\sigma_{N_2}^2}.
\end{equation}
Now, for the rate of Agent~1, from \eqref{eq:R1_gaussian} we have
\begin{align}
R_1 &\geq I(U_1;Y_1|U_2)\notag\\
&\eqtop{a} I(U_1;X,Y_1|U_2)\notag\\
&\eqtop{b} I(U_1;X|U_2) + I(U_1;Y_1|X)\notag\\
&\eqtop{c} h(X|U_2) - h(X|U_1,U_2) + r_1\notag\\
&\eqtop{d} -\frac{1}{2}\log\left(\frac{1}{2\pi e}\Big(\frac{1}{\sigma_X^2}+\frac{1-2^{-2r_2}}{\sigma_{N_2}^2}\Big) \right) \notag\\
&\qquad - \frac{1}{2}\log(2\pi eD)+ r_1\notag\\
&= \frac{1}{2}\log\frac{1}{D} - \frac{1}{2}\log\left(\frac{1}{\sigma_X^2}+\frac{1-2^{-2r_2}}{\sigma_{N_2}^2}\right)+  r_1,
\end{align}
where
\begin{enumerate}[(a)]
\item follows as $U_1-Y_1-(X,U_2)$ forms a Markov chain;
\item follows as $(U_1,Y_1)-X-U_2$ is also a Markov chain;
\item follows from the definition of $r_1$ in \eqref{eq:r_j_def};
\item follows from \eqref{eq:est_err} and \eqref{eq:r_j}.
\end{enumerate}
The rate of Agent~2 in \eqref{eq:R2_gaussian} is written in a similar way. For the sum rate, we have
\begin{align}
R_1 &+ R_2 \geq I(U_1,U_2;Y_1,Y_2)\notag\\
&\eqtop{a} I(U_1,U_2;X,Y_1,Y_2)\notag\\
&\eqtop{b} I(U_1,U_2;X) + I(U_1,U_2;Y_1,Y_2|X)\notag\\
&\eqtop{c} h(X) - h(X|U_1,U_2) + I(U_1;Y_1|X) + I(U_2;Y_2|X)\notag\\
&\eqtop{d} \frac{1}{2}\log(2\pi e\sigma_X^2) - \frac{1}{2}\log(2\pi eD)+ r_1 + r_2\notag\\
&= \frac{1}{2}\log\frac{1}{D} - \frac{1}{2}\log\frac{1}{\sigma_X^2} +  r_1+ r_2,
\end{align}
where $(a)$--$(c)$ are due to the long Markov chain $U_1-Y_1-X-Y_2-U_2$ and $(d)$ from \eqref{eq:est_err} and \eqref{eq:r_j_def}.

Using \eqref{eq:r_j_def} and the bound on $R_1$ (and $R_2$), the equivocation rate in \eqref{eq:delta1_gaussian} (and similarly \eqref{eq:delta2_gaussian}) can be written as
\begin{align}
\Delta_1  &\leq  h(X) - I(U_1;Y_1|U_2) + I(U_1;Y_1|X)\notag\\
&= \frac{1}{2}\log(2\pi e\sigma_X^2)\! -\! \frac{1}{2}\log\frac{1}{D} \! +\! \frac{1}{2}\log\left(\frac{1}{\sigma_X^2}\!+\!\frac{1-2^{-2r_2}}{\sigma_{N_2}^2}\right).
\end{align}
For sum of the equivocation rates, we have
\begin{align}
\Delta_1  +  \Delta_2  &\leq  2h(X) - I(U_1,U_2;Y_1,Y_2) + I(U_1;Y_1|X) \notag\\
&\qquad + I(U_2;Y_2|X)\notag\\
&= \log(2\pi e \sigma_X^2) - \frac{1}{2}\log\frac{1}{D} + \frac{1}{2}\log\frac{1}{\sigma_X^2}\notag\\
&=\log(2\pi e) - \frac{1}{2}\log\frac{1}{D} - \frac{1}{2}\log\frac{1}{\sigma_X^2}.
\end{align}
Finally, the bound in \eqref{eq:delta1R2_gaussian} (and similarly \eqref{eq:delta2R1_gaussian}) can be rewritten as
\begin{align}
\Delta_1 &- R_2 \notag \\
&\leq  h(X) -I(U_2;Y_2|U_1) -  I(U_1;Y_1) +  I(U_1;Y_1|X)\notag\\
&\eqtop{a} h(X) -I(U_1,U_2;Y_1,Y_2) +  I(U_1;Y_1|X)\notag\\
&= \log(2\pi e \sigma_X^2) \!-\! \frac{1}{2}\log\frac{1}{D} \! +\! \frac{1}{2}\log\frac{1}{\sigma_X^2} -r_1 - r_2 + r_1\notag\\
&= \log(2\pi e) - \frac{1}{2}\log\frac{1}{D} - r_2,
\end{align}
where $(a)$ is due to the Markov chains $U_1-Y_1-Y_2$ and $Y_1-Y_2-U_2$.\QEDA

\subsection{Converse}
Denote the messages transmitted by the agents as $J_1=f_1(Y_1^n)$ and $J_2=f_2(Y_2^n)$, and define
\begin{equation}
\label{eq:r_j_def_out}
r_j \coloneqq \frac{1}{n} I(J_j;Y_j^n|X^n).
\end{equation}
Let $\calJ$ be a subset of $\{1,2\}$ and $\calJ^c$ its complement. We have
\begin{align}
\sum_{j\in\calJ}& n(R_j+\epsilon) \geq H(J_\calJ) \notag\\
&\geq H(J_\calJ|J_{\calJ^c}) \notag\\
&= I(J_\calJ;Y^n_\calJ|J_{\calJ^c}) \notag\\
&\eqtop{a} I(J_\calJ;X^n,Y^n_\calJ|J_{\calJ^c}) \notag\\
&= I(J_\calJ;X^n|J_{\calJ^c}) + I(J_\calJ;Y^n_\calJ|X^n,J_{\calJ^c}) \notag\\
&\eqtop{b} I(J_\calJ,J_{\calJ^c};X^n) - I(J_{\calJ^c};X^n) + \sum_{j\in\calJ} I(J_j;Y^n_j|X^n) \notag\\
&\geqtop{c} \left[I(\hat{X}^n;X^n) - I(J_{\calJ^c};X^n)\right]^+ + \sum_{j\in\calJ} nr_j, \label{eq:R_out1}
\end{align}
where
\begin{enumerate}[(a)]
\item follows from the Markov chain $J_\calJ-(Y^n_\calJ,J_{\calJ^c})-X^n$;
\item follows from the Markov chain $(J_j,Y^n_j)-X^n-J_{j'}$ for $j\neq j'$;
\item follows from the data processing inequality and the definition of $r_j$ in \eqref{eq:r_j_def_out}.
\end{enumerate}
The first term in the right-hand side of \eqref{eq:R_out1} is bounded as
\begin{align}
I(\hat{X}^n;X^n) &= h(X^n) - h(X^n|\hat{X}^n)\notag \\
&= h(X^n) - h(X^n-\hat{X}^n|\hat{X}^n) \notag\\
&\geq h(X^n) - h(X^n-\hat{X}^n) \notag\\
&\geq \frac{n}{2}\log(2\pi e\sigma_X^2) - \frac{n}{2}\log(2\pi eD) \notag\\
&= \frac{n}{2}\log\frac{\sigma_X^2}{D}. \label{eq:1term}
\end{align}
We use the following lemma  to bound the second term in the right-hand side of \eqref{eq:R_out1}. The proof of the lemma is given in \cite{Prabhakaran-2004-ISIT} and \cite{Oohama-2005-TIT}.
\begin{lemma}
\label{lem:gauss}
Define $r_j \coloneqq \frac{1}{n} I(J_j;Y_j^n|X^n)$ for $j\in\{1,2\}$. Then, for $\calJ\subseteq\{1,2\}$,
\begin{align*}
\frac{1}{n}I(J_{\calJ};X^n) \leq \frac{1}{2}\log\left[\sigma_X^2\left(\frac{1}{\sigma_X^2} + \sum_{j\in\calJ}\frac{1-2^{-2r_j}}{\sigma_{N_j}^2}\right)\right].
\end{align*}
\end{lemma}
Based on Lemma~\ref{lem:gauss}, \eqref{eq:1term}, and \eqref{eq:R_out1}, we obtain
\begin{align}
\sum_{j\in\calJ} R_j &\geq \frac{1}{2}\log\frac{1}{D} - \frac{1}{2}\log\left(\frac{1}{\sigma_X^2} + \sum_{j\in\calJ^c}\frac{1-2^{-2r_j}}{\sigma_{N_j}^2}\right) \notag\\
&\qquad + \sum_{j\in\calJ} r_j.
\end{align}
Now, substituting $\calJ$ with sets $\{1\}$, $\{2\}$, $\{1,2\}$, and $\emptyset$ leads to the bounds \eqref{eq:R1_gaussian_th}--\eqref{eq:R12_gaussian_th} and \eqref{eq:D_gaussian_th} of Theorem~\ref{th:gaussian}, respectively.

The bounds on the equivocation rates are obtained as
\begin{align}
\sum_{j\in\calJ} n(\Delta_j-\epsilon) &\leq \sum_{j\in\calJ} h(X^n|J_j)\notag\\
&= \sum_{j\in\calJ} h(X^n) - H(J_j) + H(J_j|X^n) \notag\\
&\eqtop{a} \sum_{j\in\calJ} h(X^n) - H(J_j) + I(J_j;Y_j^n|X^n)  \notag\\
&\leqtop{b} \sum_{j\in\calJ} \Big[h(X^n) + I(J_j;Y_j^n|X^n)\Big] - H(J_{\calJ})\notag\\
&\leqtop{c} \sum_{j\in\calJ} \frac{n}{2}\log(2\pi e\sigma_X^2) - \frac{n}{2}\log\frac{1}{D} \notag\\
&\;\;  +  \frac{n}{2}\log\left(\frac{1}{\sigma_X^2} + \sum_{j\in\calJ^c}\frac{1-2^{-2r_j}}{\sigma_{N_j}^2}\right),
\end{align}
where
\begin{enumerate}[(a)]
\item follows since $J_j=f_j(Y^n_j)$;
\item follows since $H(J_{\calJ})\leq \sum_{j\in\calJ}H(J_j)$;
\item follows from the definition of $r_j$ in \eqref{eq:r_j_def_out} and the same techniques used to bound $H(J_{\calJ})$ in \eqref{eq:R_out1}.
\end{enumerate}
Setting $\calJ$ to $\{1\}$, $\{2\}$, and $\{1,2\}$ leads to the bounds \eqref{eq:delta1_gaussian_th}--\eqref{eq:delta12_gaussian_th}, respectively.

For the bound in \eqref{eq:delta1R2_gaussian_th}, consider
\begin{align}
n(&\Delta_1-\epsilon) \leq h(X^n|J_1)\notag\\
&= h(X^n|J_1,J_2) + I(X^n;J_2|J_1)\notag\\
&= h(X^n) - I(X^n;J_1,J_2) + H(J_2|J_1) - H(J_2|J_1,X^n)\notag\\
&\leqtop{a} h(X^n) - I(X^n;\hat{X}^n) + H(J_2) - H(J_2|X^n)\notag\\
&\leqtop{b} h(X^n) - I(X^n;\hat{X}^n) + nR_2 - I(J_2;Y_2^n|X^n)\notag\\
&\leqtop{c} \frac{n}{2}\log(2\pi e\sigma_X^2) - \frac{n}{2}\log\frac{\sigma_X^2}{D} + nR_2 - nr_2,
\end{align}
where
\begin{enumerate}[(a)]
\item follows since $J_1-X^n-J_2$ and that conditioning reduces entropy;
\item follows since $J_2=f_2(Y^n_2)$;
\item follows from \eqref{eq:1term} and the definition of $r_j$ in \eqref{eq:r_j_def_out}.
\end{enumerate}
Therefore,
\begin{equation}
\Delta_1 - R_2 \leq \frac{1}{2}\log(2\pi e) - \frac{1}{2}\log\frac{1}{D} - r_2.
\end{equation}
The bound in \eqref{eq:delta2R1_gaussian_th} can be proved similarly.\QEDA


\section{Proof of Theorem~\ref{th:gaussian_SI}: Quadratic Gaussian Case with Side Information at Eve}
\label{sec:proof_gauss_wSI}

The proof of Theorem~\ref{th:gaussian_SI} is based on a similar approach for the achievability proof of Theorem~\ref{th:gaussian} described in Appendix~\ref{sec:proof_gauss_ach}. However, in this case due to the side information at Eve, setting the auxiliary random variable $V_j$ to a constant value in Theorem~\ref{th:inner} does not always maximize the equivocation rates. %
We rewrite the bounds on the equivocation rates in Theorem~\ref{th:inner} as follows:
\begin{align}
\Delta_1  &\leq  h(X|E) - I(U_1;Y_1|U_2) + I(U_1;Y_1|X) \notag\\
&\qquad+ I(V_1;E) - I(V_1;U_2),  \label{eq:pr_delta1_gaussian_SI}\\
\Delta_2  &\leq  h(X|E) - I(U_2;Y_2|U_1) + I(U_2;Y_2|X) \notag\\
&\qquad+ I(V_2;E) - I(V_2;U_1),  \label{eq:pr_delta2_gaussian_SI}\\
\Delta_1  +  \Delta_2  &\leq  2h(X|E) - I(U_1,U_2;Y_1,Y_2) \notag\\
&\qquad + I(U_1;Y_1|X) + I(U_2;Y_2|X) \notag\\
&\qquad+ I(V_1;E) + I(V_2;E) - I(V_1;V_2), \label{eq:pr_delta12_gaussian_SI} \\
\Delta_1  - R_2  &\leq  h(X|E) -I(U_2;Y_2|U_1) -  I(U_1;Y_1) \notag\\
&\qquad+  I(U_1;Y_1|X) + I(V_1;E), \label{eq:pr_delta1R2_gaussian_SI} \\
\Delta_2 - R_1  &\leq  h(X|E) - I(U_1;Y_1|U_2) -  I(U_2;Y_2) \notag\\
&\qquad + I(U_2;Y_2|X) + I(V_2;E).  \label{eq:pr_delta2R1_gaussian_SI}
\end{align}
Then, considering the Markov chains $V_j-X-(E,U_{j'})$ for $j'\neq j\in\{1,2\}$ and the fact that the variables $X$ and $E$ as well as the auxiliary random variables $U_1$ and $U_2$ in \eqref{eq:Uj_gaussian} are Gaussian, we have four possibilities:
\begin{itemize}
\item $E$ is less noisy than $U_2$ w.r.t.\ $X$ $\Rightarrow$ $I(V_1;E)\geq I(V_1;U_2)$ $\Rightarrow$ setting $V_1=U_1$ maximizes \eqref{eq:pr_delta1_gaussian_SI};
\item $U_2$ is less noisy than $E$ w.r.t.\ $X$ $\Rightarrow$ $I(V_1;E)\leq I(V_1;U_2)$ $\Rightarrow$ setting $V_1=\emptyset$ maximizes \eqref{eq:pr_delta1_gaussian_SI};
\item $E$ is less noisy than $U_1$ w.r.t.\ $X$ $\Rightarrow$ $I(V_2;E)\geq I(V_2;U_1)$ $\Rightarrow$ setting $V_2=U_2$ maximizes \eqref{eq:pr_delta2_gaussian_SI};
\item $U_1$ is less noisy than $E$ w.r.t.\ $X$ $\Rightarrow$ $I(V_2;E)\leq I(V_2;U_1)$ $\Rightarrow$ setting $V_2=\emptyset$ maximizes \eqref{eq:pr_delta2_gaussian_SI}.
\end{itemize}
Therefore, we define
\begin{align}
T_1 &\coloneqq \max_{V_1} I(V_1;E) - I(V_1;U_2) \notag\\
&= \max\big\{0,I(U_1;E) - I(U_1;U_2)\big\},\\
T_2 &\coloneqq \max_{V_2} I(V_2;E) - I(V_2;U_1) \notag\\
&= \max\big\{0,I(U_2;E) - I(U_1;U_2)\big\},
\end{align}
and rewrite \eqref{eq:pr_delta1_gaussian_SI}--\eqref{eq:pr_delta2R1_gaussian_SI} as
\begin{align}
\Delta_1  &\leq  h(X|E) \!-\! I(U_1;Y_1|U_2) \!+\! I(U_1;Y_1|X) \!+\! T_1,  \label{eq:pr_delta1_gaussian_SI2}\\
\Delta_2  &\leq  h(X|E) \!-\! I(U_2;Y_2|U_1)  \!+\! I(U_2;Y_2|X) \!+\! T_2,  \label{eq:pr_delta2_gaussian_SI2}\\
\Delta_1  \!+\!  \Delta_2  &\leq  2h(X|E)\! -\! I(U_1,U_2;Y_1,Y_2) \notag\\
&\quad + I(U_1;Y_1|X) \!+\! I(U_2;Y_2|X) \notag\\
&\quad+ T_1 \! +\! T_2 \!+\! I(U_1;U_2)\mathbbm{1}_{\Real_{>0}}(T_1\! +\! T_2), \label{eq:pr_delta12_gaussian_SI2} \\
\Delta_1 \! - \! R_2  &\leq  h(X|E)\! -\! I(U_2;Y_2|U_1) \!-\!  I(U_1;Y_1) \notag\\
&\quad+  I(U_1;Y_1|X) \!+\! T_1 \!+\! I(U_1;U_2)\mathbbm{1}_{\Real_{>0}}(T_1), \label{eq:pr_delta1R2_gaussian_SI2} \\
\Delta_2 \!-\! R_1  &\leq  h(X|E) \!-\! I(U_1;Y_1|U_2) \!-\!  I(U_2;Y_2) \notag\\
&\quad + I(U_2;Y_2|X) \!+\! T_2 \!+\! I(U_1;U_2)\mathbbm{1}_{\Real_{>0}}(T_2),  \label{eq:pr_delta2R1_gaussian_SI2}
\end{align}
where $\mathbbm{1}_{\Real_{>0}}(\cdot)$ is the indicator function. Based on the definitions of the auxiliary random variables in \eqref{eq:Uj_gaussian} and $r_j$ in \eqref{eq:r_j}, with straightforward calculations we obtain
\begin{align}
&T_1 = \max\Bigg\{0, \frac{1}{2}\log \Bigg(1 + \frac{\frac{\sigma_{N_2}^2}{1-2^{-2r_2}}-\sigma_{N_E}^2}{\frac{\sigma_{N_1}^2}{1-2^{-2r_1}}+\sigma_{N_E}^2} \Bigg) \Bigg\}, \\
&T_2 = \max\Bigg\{0, \frac{1}{2}\log \Bigg(1 + \frac{\frac{\sigma_{N_1}^2}{1-2^{-2r_1}}-\sigma_{N_E}^2}{\frac{\sigma_{N_2}^2}{1-2^{-2r_2}}+\sigma_{N_E}^2} \Bigg) \Bigg\}, \\
&I(U_1;U_2) = \frac{1}{2}\log \Bigg( \frac{\sigma_X^2\! +\! \frac{\sigma_{N_1}^2}{1-2^{-2r_1}}}{\frac{\sigma_{N_1}^2}{1-2^{-2r_1}}\!+\!\frac{\sigma_{N_2}^2}{1-2^{-2r_2}}} \Bigg). \label{eq:IU1U2}
\end{align}

Using \eqref{eq:pr_delta1_gaussian_SI2}--\eqref{eq:IU1U2} and the proof given in Appendix~\ref{sec:proof_gauss_ach}, inequalities of Theorem~\ref{th:gaussian_SI} follows.\QEDA

\section*{Acknowledgment}
The authors thank the associate editor and anonymous reviewers for their constructive comments. They are also grateful to Dr.~Kittipong Kittichokechai for his helpful comments on an earlier draft of this work.


\bibliographystyle{IEEEtran}
\bibliography{farshad}

\end{document}